\documentstyle[12pt]{article}
 \def\be{\begin{eqnarray}}
 \def\ee{\end{eqnarray}}
 \newcommand{\bm}[1] {\mbox{\boldmath{$#1$}}}
\begin{document} 
\pagestyle{empty}
\Huge{\noindent{Istituto\\Nazionale\\Fisica\\Nucleare}}

\vspace{-3.9cm}

\Large{\rightline{Sezione di ROMA}}
\normalsize{}
\rightline{Piazzale Aldo  Moro, 2}
\rightline{I-00185 Roma, Italy}

\vspace{0.65cm}

\rightline{INFN-1214/98}
\rightline{July 1998}

\vspace{1.cm}

\large
\begin{center}{\large\bf ELECTROMAGNETIC  AND WEAK CURRENT OPERATORS FOR
INTERACTING SYSTEMS WITHIN THE  FRONT-FORM DYNAMICS}
\end{center}
\vskip 1em
\begin{center} F.M. Lev$^a$, E. Pace$^b$ and G. Salm\`e$^c$\end{center}

\noindent {\it $^a$Laboratory of Nuclear Problems, Joint Institute
for Nuclear Research, Dubna, Moscow region 141980, Russia}

\noindent{$^b$\it Dipartimento di Fisica, Universit\`a di Roma
"Tor Vergata", and Istituto Nazionale di Fisica Nucleare, Sezione
Tor Vergata, Via della Ricerca Scientifica 1, I-00133, Rome,
Italy}

\noindent{$^c$\it Istituto Nazionale di Fisica Nucleare, Sezione
di  Roma, P.le A. Moro 2, I-00185 Rome, Italy}

\begin{abstract}
Electromagnetic and weak current operators for  interacting
systems should properly
commute with the  Poincar\'e generators
and satisfy Hermiticity. The electromagnetic current  should
also satisfy ${\cal P}$  and ${\cal T}$
 covariance and continuity equation. In front-form dynamics
 the current can be  constructed
from   auxiliary operators, defined in a Breit frame where 
initial and final  three-momenta of the system are directed along the
 $z$ axis.  Poincar\'e
covariance constraints  reduce for  auxiliary operators  to the ones imposed
only by  kinematical rotations around the $z$ axis; while Hermiticity 
requires a suitable behaviour of the
auxiliary operators under rotations by $\pi$ around the $x$ or $y$ axes.
Applications to deep inelastic structure functions and electromagnetic
form factors are discussed.  Elastic and transition
form factors can be
extracted without any  ambiguity and in the elastic case  the
continuity equation is automatically satisfied, once Poincar\'e,
${\cal P}$  and ${\cal T}$ covariance, together with  Hermiticity,
are imposed.

\end{abstract}

{\bf{Pacs:}} 11.40-q 11.40.Dw 13.40.Gp 13.60.Hb 25.30.Bf

\vspace{1.cm}
\hrule width5cm
\vspace{.2cm}
\noindent{\normalsize{ Submitted to {\bf Nucl. Phys. A}}}

\newpage
\pagestyle{plain}
\section{Introduction}
\label{S1}

   Experiments on modern accelerators make it possible to
investigate a variety of electromagnetic (em) and weak properties of
hadrons. A comprehensive theoretical analysis of these
properties encounters serious difficulties, since perturbative QCD
does not apply to the bound state problem. In view  of these
difficulties, effective models have been developed, but both the
theory and models use the current operator  as a fundamental input
for evaluating elastic and
inelastic form factors of relativistic interacting systems. For this
reason, it is important to understand which constraints on the
current operators can be imposed taking into account only general
properties, e. g.,  Poincar\'e covariance and  Hermiticity.

For instance, in the relevant case of deep inelastic scattering
(DIS)  the cross-section
is fully defined by the hadronic tensor
\begin{equation}
W^{\mu\nu}=\frac{1}{4\pi}\int\nolimits e^{\imath qx} \langle P',\chi'|
J^{\mu}(x)J^{\nu}(0) |P',\chi'\rangle d^4x
\label{01}
\end{equation}
where $q$ is the momentum transfer and
$|P',\chi'\rangle $ is the initial state of the nucleon
with  four-momentum $P'$ and  internal wave function
$\chi'$.
This tensor will have correct transformation properties relative
to the Poincar\'e group (i.e., $W^{\mu\nu}$
will be a true tensor) only if both  the state
$|P',\chi'\rangle$ and the operator
$J^{\mu}(x)$ have  correct transformation properties with respect to
the {\em same} representation of the Poincar\'e group. In
 the parton model proposed by
Bjorken \cite{Bjor} and Feynman \cite{Feyn}  the nucleon is described
as a bound system, while the (em or weak) current operator
$J^{\mu}(x)$ ($x$ is a point in Minkowski space) is taken in impulse
approximation (IA), i.e., it is the same as for noninteracting particles.
According to the present theory based on the operator product
expansion \cite{Wil} and factorization theorem \cite{Sterman},
the parton model (even in the Bjorken limit) is accurate  up
to anomalous dimensions and perturbative QCD corrections;  therefore
the corrections to the parton model can be considered in the framework
of perturbation theory. However the nucleon is a  bound state of quarks
and gluons and cannot be described perturbatively.  Therefore, in
principle, the problem arises  whether  the
operator $J^{\mu}(x)$, treated perturbatively at large $Q=|q^2|^{1/2}$,
is compatible with the
correct transformation properties of  bound states.

In the case of model approaches,  another example of the necessity of
compatibility between the generators of the Poincar\'e group and the
current operators is clearly met in  the investigation of elastic and
inelastic hadron form factors within the  front-form Hamiltonian
dynamics \cite{Dir}. In this  framework
   hadron form factors have generally
been calculated assuming  that, in the reference frame where
$q^+=0$, the component $J^+(0)$ can be  taken in IA (the $\pm$
components of
four-vectors are defined as $p^{\pm}=(p^0+p^z)/\sqrt{2}$). The main
argument in favor of this assumption
(see ,e.g., Refs. \cite{Wein,Close,LepBr,Namysl}) is that in the
reference frame where $q^+=0$
the production of pairs from the vacuum is forbidden by
momentum conservation and the operator $J^+(0)$ gives a contribution
only for positive-energy components of Dirac spinors. However, the
hadron form factors are
determined by matrix elements of $J^{\mu}(0)$ between   initial
and final bound states and, by analogy with the case of DIS, the problem
arises whether the assumption that $J^+(0)$ is free is compatible
with the correct transformation properties of its matrix elements.
Indeed, consider for example the elastic electron-deuteron
scattering in the Breit frame of the deuteron, i.e., in the
reference frame where the initial and final three-momenta ${\bm P}'$ and
${\bm P}"$ satisfy the condition ${\bm P}'+{\bm P}"=0$. If
$\lambda'$ and $\lambda"$ are the deuteron helicities in the
initial and final states, respectively, and
$I_{\lambda"\lambda'}=\langle \lambda"|J^+(0)|\lambda'\rangle $
then, as follows from ${\cal P}$ and ${\cal T}$ covariance, all the
matrix elements
$I_{\lambda"\lambda'}$ can be expressed in terms of $I_{11}$,
$I_{00}$, $I_{10}$ and $I_{1,-1}$. As follows from
Poincar\'e covariance, current conservation and Hermiticity, the
elastic electron-deuteron scattering is described by three
independent real form factors and therefore the above matrix elements
are not independent. As shown in Refs. \cite{GrKo,CCKP} and others,
if $\eta=Q^2/4m_d^2$, with  $m_d$  the deuteron mass, then the following
constraint, called " angular condition" must be fulfilled in the $q^+=0$
 frame, viz.
\begin{equation}
(1+2\eta)I_{11}-I_{00}-(8\eta)^{1/2}I_{10}+I_{1,-1}=0.
\label{1}
\end{equation}
However this relation is not satisfied if the matrix elements
$I_{\lambda"\lambda'}$  are calculated with the free operator,
$J_{free}^+(0)$, and therefore interactions term are needed.

 A way to avoid  this difficulty has been proposed, e.g., in
\cite{GrKo}, where it is noted
that the three form factors can be determined by using the free
operator $J_{free}^+(0)$ for
calculating only  three
matrix elements, while the fourth one can be
determined (if necessary) from Eq. (\ref{1}). It is clear that such
a procedure contains a large extent of freedom.
 In absence of any dynamical scheme, only the comparison of the
results with the data can yield insight, if any, on the
 choice of the three
matrix elements to be preferred.
Another approach has been proposed in
\cite{KS} within  the covariant formulation of the
front-form dynamics. In this approach the   matrix elements
$\langle \lambda"|J^{\mu}(0)|\lambda'\rangle $
with  $J^{\mu}(0) \equiv J_{free}^{\mu}(0)$ are given by the sum  of
eleven contributions. Only three of them  depend upon the physical
form factors (as they must do if the operator fulfills the Poincar\'e
covariance and the current conservation), while the other
contributions contain the null vector  $\omega^{\mu}$, which
determines the direction of the null plane in Minkowski space, and
are unphysical.  The physical form factors can be formally
obtained if, following Ref. \cite{Gl}, two matrix elements are
calculated by using $J_{free}^+(0)$ and the third matrix element is
calculated by using $J_{free}^j(0)$ with $j=1$ or $j=2$.  However, it
remains unclear whether the contribution of interaction terms in the
current operator to the physical form factors is important.

   In view of the above discussion it is important to know the
constraints imposed by Poincar\'e covariance  on the exact current
operator $J^{\mu}(x)$,   and, as far as the em current is concerned
the further constraints imposed by current conservation, parity and
time reversal.
These constraints were investigated in detail in Refs.
\cite{PK,lev0,lev}. As shown in Ref. \cite{PK}, if the mass and
spin operators for the system as a whole are diagonalized, the
matrix elements of the current operator can be expressed in terms
of reduced matrix elements which are not constrained by Poincar\'e
covariance.  However for
practical calculations it is desirable to know all the constraints
directly in terms of operators (see, e.g., \cite{lev}), and
furthermore the Hermiticity condition (discussed in detail in this
paper) and the cluster separability (see, e.g., Refs.
\cite{sok1,sok,CP,Mutze}) are to be satisfied.  For systems
with a fixed number of  interacting relativistic particles a
current operator satisfying Poincar\'e covariance, Hermiticity and cluster
separability to order $(v/c)^4$ has been constructed in Ref.
\cite{lev0}. An exact solution in the point form of dynamics
has been considered in Ref. \cite{lev}, where the current operator
is expressed in terms of auxiliary operators  defined in the
equal-velocity frame, with the $z$ axis directed along the three-momentum
transfer. In particular,  it has been shown that the
free-current operator  represents a possible solution for the
auxiliary operators.

  Finally, we  mention that the  requirement of locality,  in
the sense that the commutator $[J^{\mu}(x),J^{\mu}(y)]$
should  vanish when $x-y$ is a space-like vector,
could be used for choosing between different solutions, if they exist,
 satisfying Poincar\'e covariance and the other general properties.

  Aim of this paper is the investigation of the  constraints imposed
on the current operator by
extended Poincar\'e covariance (continuous + discrete
transformations), Hermiticity and
current conservation within the front-form dynamics \cite{Dir}. This
dynamics is widely adopted   and exhibits many interesting features,
such as the largest set of kinematical Poincar\'e  generators and
the boundedness from below of the $P^+$ component (see, e.g.,
\cite{Namysl,KP}). We have followed an approach analogous to the one
of Ref. \cite{lev}, but applied directly in the front form. In Ref.
\cite{lev}, by using the unitary equivalence of the different forms
of relativistic dynamics \cite{SoSh}, the current operators in the front
and the instant form corresponding to the particular solution found in the
point form, have been constructed. It should be pointed out that   the
point-form auxiliary operators obtained from the free current   generate
both non interacting and interacting terms in the corresponding front-form
(instant) operators. Therefore it is interesting to investigate if , also
in the front form,   auxiliary operators obtained directly from a free
current can produce an exact solution. In Ref. \cite{lev} the direct
construction of the current in the front-form was not considered,
arguing the presence of extra difficulties with respect to the point form.
In this paper such  problems will be solved expressing the current operator
in terms of auxiliary operators, which act only through internal
variables and depend upon  given masses for the initial and final system
(namely considering a spectral decomposition of the current operator).
A particular attention has to be devoted to the choice of
the reference frame where the auxiliary operators are defined. As a
matter of fact, in the conventional (instant form) approach  to
elastic scattering the form factors are generally evaluated in the
reference frame where $q^0=0$ and then, as follows
from rotational invariance of the ordinary three-dimensional space,
the $z$ axis can be chosen along the common direction of the initial and
final  three-momenta of the system. Hence the problem becomes
symmetric with respect  to rotations
around the $z$ axis ($\equiv$ the spin quantization axis). In the
front form, the reference frame analogous to the one adopted in the
conventional approach is the frame  where  $q^+=0$ (for both elastic
and inelastic scattering). In this case, it is again
possible to find a reference  frame where  both  initial and final
three-momenta of the system  are directed along the same vector
${\bm n}$, but
${\bm n}$ obviously cannot coincide with the $z$ axis. Therefore the
rotational invariance around the $z$ axis is lost.  It is worth
noting that in the front form the rotation around the $z$ axis are
kinematical, while those around the $x$ and $y$ axes are dynamical.
In order to take advantage of this peculiarity of the $z$ axis and
to restore the symmetry of the physical problem, we perform our
analysis in the Breit frame where the initial and final
three-momenta of the system are
directed along the $z$  axis, and as a consequence $q^+ \not = 0$.

One of the main results of our investigation is  that,  in order to
satisfy the    Poincar\'e covariance,
the auxiliary  operators in our  Breit frame have to be covariant
only with respect to   rotations around the $z$ axis.

The paper is organized as follows: in Sects. \ref{S0}-\ref{S5} the
general formalism relevant for relativistic interacting systems is
 presented; in Sects. \ref{S6} and \ref{S7} the constraints imposed on
the current operators are explicitly given; in Sect. \ref{S8} the
matrix elements of the current are discussed;
in Sect. \ref{S9} the application to
DIS is investigated; in Sect. \ref{S10} the cases of elastic 
and inelastic scattering are considered. Finally in Sect. 
\ref{S11} conclusions are drawn.

\section{General Formalism}
\label{S0}

 Let $P$ be the operator of the four-momentum for the system under
consideration and
$M^{\mu\nu}$ ($M^{\mu\nu}=-M^{\nu\mu}$) be the representation
generators of the Lorentz group.
We shall always assume that the commutation relations for the
representation generators of the Poincar\'e group are realized in the form
$$[P^{\mu},P^{\nu}]=0, \quad [M^{\mu\nu},
P^{\rho}]= -\imath({\eta}^{\mu\rho}P^{\nu}-{\eta}^{\nu\rho}P^{\mu}),$$
\begin{equation}
[M^{\mu\nu},M^{\rho\sigma}]=-\imath ({\eta}^{\mu\rho}
M^{\nu\sigma}+{\eta}^{\nu \sigma}M^{\mu\rho}-{\eta}^
{\mu\sigma}M^{\nu\rho}-{\eta}^{\nu\rho}M^{\mu\sigma})
\label{02}
\end{equation}
where $\mu,\nu,\rho,\sigma=0,1,2,3$, the metric tensor in Minkowski
space has the nonzero components $\eta^{00}=-\eta^{11}=-\eta^{22}=
-\eta^{33}=1$, and we use the system of units with $\hbar=c=1$.

\begin{sloppypar}
 As explained in  well-known textbooks and monographs (see, e.g.,
Refs. \cite{AB,BLOT}), matrix elements of field operators  have the
correct transformation properties relative to
transformations of the Poincar\'e group, only if the transformation
of the operators are compatible with the transformations of the
states. This implies that the current operator should satisfy the
conditions
\begin{equation}
exp(\imath Px) J^{\mu}(0)exp(-\imath Px)=J^{\mu}(x),
\label{31}
\end{equation}
\begin{equation}
U(l)^{-1}J^{\mu}(x)U(l)=L(l)^{\mu}_{\nu}J^{\nu}(L(l)^{-1}x)
\label{32}
\end{equation}
where $L(l)$ is the element of the Lorentz group corresponding to
$l\in SL(2,C)$
and $U(l)$ is the representation operator corresponding to $l$.
\end{sloppypar}

In particular
\begin{equation}
U(l)^{-1}J^{\mu}(0)U(l)=L(l)^{\mu}_{\nu}J^{\nu}(0)
\label{33}
\end{equation}
Therefore, as a consequence of Lorentz
covariance, one has
\begin{equation}
[M^{\mu\nu},J^{\rho}(0)]= -\imath(\eta^{\mu\rho}J^{\nu}(0)-\eta^{\nu\rho}
J^{\mu}(0))
\label{34}
\end{equation}

\begin{sloppypar}
Since some of the Poincar\'e group generators describing Poincar\'e
transformations of the bound states are necessarily interaction
dependent, the above expressions show that $J^{\mu}(x)$ is a
relativistic vector operator only if it
depends on the interaction present in the system under consideration.
In general,
it is not clear whether this condition can be compatible with the usual
assumption that $J^+(0)$ is free. Moreover, as already explained,
there exist cases where this assumption is definitely incompatible with
Poincar\'e covariance and current conservation (see, e.g., Refs.
\cite{GrKo,KoSt}).
\end{sloppypar}

\begin{sloppypar}
\section{Irreducible representations of the Poincar\'e group with
positive mass}
\label{S2}
\end{sloppypar}

  In order to describe a relativistic system of interacting particles it is
necessary to choose first an explicit form of the unitary irreducible
representation (UIR) of the Poincar\'e group describing an elementary
particle of mass $m>0$ and spin $\bm{s}$. There are many equivalent ways to
construct an explicit realization of such a representation \cite{Wig,Nov}.
We choose the realization which is convenient in the front-form dynamics
\cite{Dir}.

 Let $p$ be the particle 4-momentum, $g=p/m$ be the particle 4-velocity
and ${\bm s}$ be the spin operator. Since $p^2=m^2$, only three
components of $p$ are independent. We choose ${\bm p}_{\bot}$ and
$p^+$ as such components, where ${\bm p}_{\bot}\equiv (p_x,p_y)$.
Let $\sigma$ be the projection of the spin on the $z$ axis.  The
one-particle space can be chosen as the
space of functions $\phi(p,\sigma)=\phi({\bm p}_{\bot},p^+,\sigma)$
with the norm
\begin{equation}
\langle\varphi | \varphi\rangle=\sum_{\sigma}\int\nolimits |\phi({\bm
p}_{\bot},p^+,\sigma)|^2d\rho({\bm p}_{\bot},p^+)
\label{2}
\end{equation}
where
\begin{equation}
d\rho({\bm p}_{\bot},p^+)=\frac{d^2{\bm p}_{\bot}dp^+}{2(2\pi)^3p^+}
\label{3}
\end{equation}

If an element of the Poincar\'e group $(a,l)$ is defined by the
four-vector $a$ and by the matrix $l\in SL(2,C)$, then the
corresponding representation operator acts as
\begin{equation}
\langle p,\sigma|U(a,l)|\phi \rangle = U(a,l)\phi(p,\sigma)=exp(\imath p'a)
\sum_{\sigma'} D_{\sigma\sigma'}^s[W(l,g')]\phi(p',\sigma')
\label{5}
\end{equation}
where $p'=L(l)^{-1}p$ and $W(l,g')$ is the front-form Wigner rotation
defined as
\begin{equation}
W(l,g')=\beta(g)^{-1}l\beta(g')
\label{Wigner}
\end{equation}
The matrices $\beta(g)\in$ SL(2,C) represent  the
front-form boosts and their components are given by
\begin{equation}
\beta_{11}=\beta_{22}^{-1}=2^{1/4}(g^+)^{1/2},\quad \beta_{12}=0,
\quad \beta_{21}=(g_x+\imath g_y)\beta_{22}
\label{4}
\end{equation}
In Eq. (\ref{5}), $D^s(u)$ is the matrix of the UIR
of the group $SU(2)$ with the spin $s$, corresponding to
$u\in SU(2)$ (it is easy to verify that $W(l,g')\in SU(2)$).

 A direct calculation shows that, for the UIR defined
by Eqs. (\ref{5}) and (\ref{Wigner}), the generators have the
well-known form (see, for example, Refs. \cite{BarHal,Ter})
\begin{eqnarray}
&&P^{+} = p^{+},\quad{\bm P}_{\bot}={\bm p}_{\bot},\quad P^-=p^-=
\frac{m^2+ {\bm p}_{\bot}^2}{2p^+}, \nonumber\\
&&M^{+-} = \imath p^+\frac{\partial}{\partial p^+},\quad
M^{+j}=-\imath p^+\frac{\partial}{\partial
p^j},\quad M^{xy}={\ell}^z({\bm p}_{\bot})+s^z, \nonumber\\
&&M^{-j}=-\imath(p^j\frac{\partial}{\partial p^+}+
p^-\frac{\partial}
{\partial p^j})-\frac{\epsilon_{jl}}{p^+}(ms^l+p^ls^z)
\label{6}
\end{eqnarray}
where a sum over $j,l=x,y$ is assumed, $\epsilon_{jl}$ has the
components $\epsilon_{xy}=-\epsilon_{yx}=1, \,
\epsilon_{xx}=\epsilon_{yy}=0$ and
${\bm {\ell}}({\bm p})=-\imath {\bm p}\times (\partial / \partial {\bm p})$
is the orbital angular-momentum operator.

 The presence of  the matrices $\beta(g)$ in Eq. (\ref{Wigner})
is very relevant. Let $B$ be a subgroup of SL(2,C) such that $b\in B$
if $b_{11}=b_{22}^{-1}>0$, $b_{12}=0$ and $b_{21}$ is an arbitrary
complex number. Then, it is clear
from Eq. (\ref{4}) that $ B$ is the set of the front-form boosts and
one can verify by a direct
calculation that
\begin{equation}
b\beta(g)=\beta(L(b)g)\quad  \mbox{if} \quad b\in B
\label{9}
\end{equation}
Therefore, as follows from Eqs. (\ref{Wigner}) and (\ref{9}), the Wigner
rotations corresponding to $b\in B$ are equal to 1 and, as follows from
Eq. (\ref{5}), the
action  of the representation operators corresponding to $b\in B$ and
$a=0$ is especially
simple, viz.
\begin{equation}
U(b)\phi(p,\sigma)=\phi(L(b)^{-1}p,\sigma)
\label{10}
\end{equation}
The representation generators of the group $B$ are $M^{+-}$ and
$M^{+j}$ (see, e.g., \cite{KP,riv}) and it is clear from Eq. (\ref{10})
why they do not depend
on ${\bm s}$ (see Eq. (\ref{6})). The important role of the group
property (\ref{9}) has been pointed out in Ref. \cite{KoTer}.

 Each element of the group SL(2,C) can be uniquely written as
$l=\beta(g)u$, where $u\in SU(2)$ (see, e.g., Ref. \cite{Nov}). Another
possible representation is $l=\alpha(g)u'$  \cite{Nov}, where
\begin{equation}
\alpha(g)=\frac{g^0+1+{\bm {\tau } }{\bm g}}
{[2(g^0+1)]^{1/2}}
\label{11}
\end{equation}
and $\bm{\tau} $ are the Pauli matrices. The matrices
$\alpha(g)$ represent the instant-form boosts and do not form a group.
The choice of $\alpha(g)$ instead of $\beta(g)$ is convenient for
investigating discrete symmetries and conventional
three-dimensional rotations. In particular one has
\begin{equation}
u\alpha(g)=\alpha(L(u)g)u \label{12} \end{equation} The relation between the
matrices $\alpha(g)$ and $\beta(g)$ is \begin{equation}
\beta(g)=\alpha(g)v(g), \quad v(g)=\frac{1+g^0+g^3+\imath
\epsilon_{jl}\tau^jg^l}{[2(1+g^0)(g^0+g^3)]^{1/2}}
\label{13}
\end{equation}
where $v(g) \in SU(2)$ is the so called Melosh matrix \cite{Mel},
which in the given context was first considered in Ref. \cite{Ter}.
In particular, note that if $g^l=0$ ($l=1,2$) then  $v=1$. Such a
property will be used in the following sections.

 Let us now consider the discrete symmetries, space reflection and time
reversal. For the reasons which
will be clear later, it is convenient  to consider not the
full space reflection ${\cal P}$, but only the reflection relative the
$y$ axis, ${\cal P}_y={\cal P} R_y(\pi)$ (where $R_y(\pi)$ is a rotation around
 the $y$ axis by  $\pi$). The action of the corresponding operator is
given by
\begin{equation}
\langle p|U_y|\varphi \rangle =U_y \varphi (p)=\eta_{\cal P}
exp(-\imath\pi s_y)\varphi ({\tilde
p}) \label{7}
\end{equation}
where $\eta_{\cal P}$ is the ${\cal P}$ parity of the particle under
 consideration, ${\tilde p}$ differs from $p$ by the sign of the
 $y$ component: ${\tilde p}=(p_x,-p_y,p^+,p^-)$ and the action of $R_y(\pi)$
 can be obtained  from Eq. (\ref{5}).

 Instead of ${\cal T}$ covariance we will consider $\theta$ covariance, where
$\theta ={\cal {P T}}$. The action of the corresponding antiunitary operator
is given by
\begin{equation}
\langle p,\sigma|U_{\theta}|\varphi \rangle= U_{\theta}\varphi(p,
\sigma)=\eta_{\theta} \overline{\langle p,\sigma| exp(-\imath\pi s_y)
|\varphi \rangle}
\label{8}
\end{equation}
where $\eta_{\theta}$ is the $\theta$ parity and the bar means the
complex conjugation.

\begin{sloppypar}
\section{Representations of the extended Poincar\'e group for
systems of noninteracting particles}
\label{S3}
\end{sloppypar}

 The space ${\cal H}$ for the representation of the
Poincar\'e group describing a system of $N$ free
particles with the masses $m_i$ and spins $s_i$ $(i=1,2,...,N)$ can be
realized as the space of functions
$\phi({\bm p}_{1\bot},p_1^+,\sigma_1,...,
{\bm p}_{N\bot},p_N^+,\sigma_N)$ with the norm
\begin{equation}
\langle \varphi|\varphi\rangle=\sum_{\sigma_1...\sigma_N}\int\nolimits
|\phi({\bm
p}_{1\bot}, p_1^+,\sigma_1,...,{\bm p}_{N\bot},p_N^+,\sigma_N)|^2
\prod_{i=1}^{N} d\rho({\bm p}_{i\bot},p_i^+)
\label{14}
\end{equation}

  Instead of the variables ${\bm p}_{1\bot}$, $p_1^+$,...,
${\bm p}_{N\bot}$, $p_N^+$, we introduce the variables ${\bm P}_{\bot}$,
$P^+$, ${\bm k}_1$,...,${\bm k}_N$, where $P=p_1+...+p_N$ is the total
four-momentum, and ${\bm k}_i$
is the spatial part of the four-vector
\begin{equation}
k_i=L[\beta(G)]^{-1}p_i,
\label{15}
\end{equation}
with  $G=P/M_0$ and $M_0=|P|\equiv |P^2|^{1/2}$. The action of the boost
$L[\beta(G)]^{-1}$ is such that $P^{'}=L[\beta(G)]^{-1}P
\equiv({\bm P}^{'}_{\bot}=0, P^{'+}=M_0/\sqrt{2},P^{'-}=M_0/\sqrt{2})$.

As follows from Eqs. (\ref{4}) and (\ref{15}), it is also possible
to use the following internal variables:
\begin{equation}
\xi_i=\frac{p_i^+}{P^+}=\sqrt{2} \frac{k_i^+}{M_0},\quad {\bm k}_{i\bot}=
{\bm p}_{i\bot}-\xi_i{\bm P}_{\bot}
\label{16}
\end{equation}

\begin{sloppypar}
The four-vectors $p_i$ have canonical components $(\omega_i({\bm p}_i),
{\bm p}_i)$, and the four-vectors $k_i$ have the components
$(\omega_i({\bm k}_i),{\bm k}_i)$, where $\omega_i({\bm k})=(m_i^2+
{\bm k}^2)^{1/2}$ and $k_z=(\xi-1/2)M_0$. In turn, only $N-1$ vectors
${\bm k}_i$ are
independent since, as follows from Eqs. (\ref{15}) and (\ref{16}),
${\bm k}_1+...+{\bm k}_N=0$, i.e., ${\bm k}_i$ are intrinsic three-momenta.
It is easy to show that $M_0=\omega_1({\bm k}_1)+...+\omega_N({\bm k}_N)$.
\end{sloppypar}

\begin{sloppypar}
A direct calculation shows that
\begin{eqnarray}
&&\prod_{i=1}^{N}d\rho({\bm p}_{i\bot},p_i^+)=
d\rho({\bm P}_{\bot},P^+)d\rho (int),\nonumber\\
&& d\rho (int)=2(2\pi)^3M_0
\delta^{(3)}({\bm k}_1+\cdots +{\bm k}_N)
 \prod_{i=1}^{N}d\rho_i({\bm k}_{i\bot},k_i^+)
\label{17}
\end{eqnarray}
Therefore the space ${\cal H}$ can be realized
as the space of functions
$\phi ({\bm P}_{\bot},P^+;{\bm k}_1,\sigma_1,...,{\bm k}_N,\sigma_N )$
 such that

\begin{equation}
\langle \phi|\phi\rangle=\sum_{\sigma_1...\sigma_N}\int\nolimits
|\phi({\bm P}_{\bot},P^+;{\bm k}_1,\sigma_1,...,{\bm k}_N,\sigma_N)|^2
d\rho({\bm P}_{\bot},P^+)d\rho (int)
\label{19}
\end{equation}

Let us also define the "internal" space ${\cal H}_{int}$ as the space of
functions $\chi({\bm k}_1,\sigma_1,...,{\bm k}_N,\sigma_N)$
such that the norm is equal to
\begin{equation}
\langle \chi|\chi\rangle=\sum_{\sigma_1...\sigma_N}\int\nolimits
|\chi({\bm k}_1,\sigma_1,...,{\bm k}_N,\sigma_N)|^2
d\rho (int)
\label{18}
\end{equation}

\end{sloppypar}

Note that in front-form dynamics the operators
${\bm P}_{\bot}$ and $P^+$ are always equal to the operators of
multiplication by the corresponding variables. Therefore the
use of the same notations (${\bm P}_{\bot},P^+$) for both the
variables and the operators should not lead to misunderstanding.
Since the structure of the operators (${\bm P}_{\bot},P^+$) is
clear, in the following we will consider only the structure of
the remaining seven generators of the Poincar\'e group.
For noninteracting particles they are equal to
sums of the corresponding one-particle generators given by Eq. (\ref{6}). A
direct calculation of these sums shows that in the
variables ${\bm P}_{\bot},P^+,{\bm k}_1,...,{\bm k}_N$ one has

\begin{eqnarray}
&&P^-=\frac{M_0^2+{\bm P}_{\bot}^2}{2P^+},\quad
M^{+-}=\imath P^+\frac{\partial}{\partial P^+},\nonumber\\
&&M^{+j}=-\imath P^+\frac{\partial}{\partial P^j},\quad
M^{xy}={\ell}^z({\bm P}_{\bot})+S_0^z,\nonumber\\
&& M^{-j}=-\imath(P^j\frac{\partial}{\partial P^+}+
P^-\frac{\partial}{\partial P^j})-
\frac{\epsilon_{jl}}{P^+}(M_0S_0^l+P^lS_0^z)
\label{20}
\end{eqnarray}
where ${\bm {\ell}}({\bm P})=-\imath {\bm P}\times
(\partial / \partial {\bm P})$.

 The operator ${\bm S}_0$ in Eq. (\ref{20}) is the spin operator
for the system as a whole. It acts only through the
variables of the space ${\cal H}_{int}$ and is unitarily equivalent
to the spin operator in the conventional form
(see, e.g., Refs. \cite{BarHal,Ter,BKT,riv}):
\begin{equation}
{\bm S}_0= \{\prod_{i=1}^{N} D^{s_i}[v(\frac{k_i}{m_i})]\}^{-1}
({\bm {\cal L}}+{\bm s}_1+...+{\bm s}_N)
\{\prod_{i=1}^{N} D^{s_i}[v(\frac{k_i}{m_i})]\}
\label{21}
\end{equation}
where ${\bm {\cal L}}$ is the total internal orbital angular momentum
operator.

We see that the many-particle generators have the same form as the
free ones, if in Eq. (\ref{6}) $p$ is replaced by $P$, $m$ by $M_0$ and
${\bm s}$ by ${\bm S}_0$.

 It is possible to show that the same is valid for the many-particle
operator $U_y$. However, for practical purposes it is sufficient to
represent the action of this operator in the form
\begin{eqnarray}
&&\langle P_x,P_y,P^+|U_y|\varphi \rangle = U_{y,int}
\langle P_x,-P_y,P^+|\varphi \rangle
\label{22}
\end{eqnarray}
where, as follows from Eqs. (\ref{7}) and (\ref{16}), the action
of the operator $U_{y,int}$ in the space of internal variables is
given by
\begin{equation}
\langle{\bm k}_1,...,{\bm k}_N |U_{y,int}| \chi \rangle =
\{\prod_{i=1}^{N} \eta_{{\cal P}_i}exp(-\imath\pi
s_{iy})\} \chi ({\bm {\tilde{k}}}_1,...,{\bm {\tilde{k}}_N)
\label{23}}
\end{equation}
where $\eta_{{\cal P}_i}$ is the internal ${\cal P}$ parity of
particle $i$, and ${\bm {\tilde{k}}}_i \equiv (k_{ix},-k_{iy},k_{iz})$.

 Analogously, as follows from Eq. (\ref{8}), the action of the operator
$U_{\theta}$ for the system as a whole can be written as
\begin{eqnarray}
&&\langle {\bm P}_{\bot},P^+; S S_z|U_{\theta}|\varphi \rangle =
 \eta_{\theta} \overline{\langle {\bm P}_{\bot},P^+; S S_z| exp(-\imath\pi
S_{0y})\} |{\varphi}\rangle}
\label{25}
\end{eqnarray}

\begin{sloppypar}
\section{Systems of interacting particles in the front form of
dynamics}
\label{S4}
\end{sloppypar}

\begin{sloppypar}
 If the particles interact with each other, then the representation
space remains the same as in the case of  free particles, but
the representation generators of the Poincar\'e group
differ from the corresponding free generators. One of the simplest way to
preserve the relativistic commutation relations is to replace $M_0$ in
Eq. (\ref{20}) by a mass operator ${\cal M}$ which acts only
through the variables of the space ${\cal H}_{int}$ and
commutes with ${\bm S}_0$. Then
\begin{eqnarray}
&&P^-=\frac{{\cal M}^2+{\bm P}_{\bot}^2}{2P^+}, \quad
M^{+-}=\imath P^+\frac{\partial}{\partial P^+},\nonumber\\
&&M^{+j}=-\imath P^+\frac{\partial}{\partial P^j},\quad
M^{xy}=l^z({\bm P}_{\bot})+S_0^z,\nonumber\\
&& M^{-j}=-\imath(P^j\frac{\partial}{\partial P^+}+
P^-\frac{\partial}{\partial P^j})-
\frac{\epsilon_{jl}}{P^+}({\cal M}S_0^l+P^lS_0^z)
\label{26}
\end{eqnarray}

\indent Such a procedure was first proposed by Bakamjian and
Thomas \cite{BT}.
According to the Dirac classification \cite{Dir}, the generators in
Eq. (\ref{26}) are given in the front form of dynamics, since this form is
characterized by the condition that only the operators $P^-$
and $M^{-j}$ are interaction dependent, while all the other
seven generators are free.
\end{sloppypar}
 In this procedure, however,  cluster
separability \cite{sok1,sok,CP,Mutze} is not implemented.
In order to satisfy cluster
separability the spin operator in
the general case ($N \geq 3$) has to be interaction dependent and the
generators can be obtained
from Eq. (\ref{26}) by replacing ${\cal M }$ by
$M=A{\cal M}A^{-1}$ and ${\bm S}_0$
by the operator ${\bm S}$, such that $S^z=S_0^z$ and $A{\bm S}_0A^{-1}=
{\bm S}$ \cite{lev2,KP,riv}, where the unitary operator $A$ acts only
through the internal variables. In this case the operators $M$ and
${\bm S}$ must commute  with each other.  The operator $A$ is the
front-form analog of the
Sokolov packing operator \cite{sok}; an explicit expression for $A$ can
be found, for example, in Refs. \cite{CP,KP,riv,lev2}. The choice $A=1$ is
possible  when
in a system of $N$ particles there exists only the $N$-particle
interaction or there is a confining interaction (see, e.g.,
\cite{Mutze,KP}).

 It is important to note that the form of the operators in Eq.
(\ref{26}) (even if ${\bm S}_0$ is replaced by ${\bm S}$) does not
explicitly depend on the number of particles. Therefore one could argue
that such a form can also be valid when
the number of particles is infinite and even when this number is not a
conserved physical quantity. Indeed, according to our intuition,
there should always exist a representation in which the external
motion is purely kinematical, while all the information about
dynamics is contained in the mass operator which acts only
through internal variables. The representation (\ref{26}) has
just such properties and therefore one might expect that any other
representation in the front form is unitarily equivalent to that
given in Eq. (\ref{26}).

 A difficulty in front-form dynamics is that the
operators $U_{\cal P}$ and $U_{\cal T}$ corresponding to space
reflection and
time reversal should necessarily be interaction dependent. This follows in
particular from the relations
\begin{equation}
U_{\cal P}P^+U_{\cal P}^{-1}=U_{\cal T}P^+U_{\cal T}^{-1}=P^-
\label{27}
\end{equation}
However, as noted by
Coester \cite{Coes}, the discrete transformation ${\cal P}_y$ such that
${\cal P}_y\, x:= \{x^0,x_1,-x_2,x_3\}$ leaves the light cone $x^+=0$
invariant, and therefore it is kinematical.
The full space reflection ${\cal P}$ is the product of ${\cal P}_y$ and a
dynamical rotation around the $y$ axis by $\pi$. Thus ${\cal P}$ is not
an independent dynamical
transformation to be considered besides the rotations around transverse axes.
Similarly the transformation $\theta$ leaves $x^+=0$ invariant and
${\cal T}=\theta {\cal P}_yR_y(\pi)$. Therefore
the interaction dependence of the operators
$U_{\cal P}$ and $U_{\cal T}$ in the front form does not mean that there
are discrete dynamical symmetries in addition to the rotations around
transverse axes. We conclude that the operators $U_y$ and
$U_{\theta}$ are interaction independent and can be chosen to be the
same as for a system of free
particles (see the preceding section). Then the generators given by
Eq. (\ref{26}) will
satisfy extended Poincar\'e covariance if
\begin{eqnarray}
&&U_{y,int}MU_{y,int}^{-1}=M,\quad
U_{y,int}AU_{y,int}^{-1}=A,\nonumber\\
&&U_{\theta}MU_{\theta}^{-1}=M,\quad U_{\theta}AU_{\theta}^{-1}=A
\label{28}
\end{eqnarray}

  Let $\Pi_i$ be the orthogonal projector onto the subspace
${\cal H}_i\equiv \Pi_i{\cal H}$ corresponding to the
eigenvalue of the operator $M$ equal to $M_i$ and to the eigenvalue
of the spin operator equal to $S_i$. Therefore by
analogy with Ref. \cite{PK} we work in the representation where the
mass and spin operators are diagonalized. In constituent quark models
the spectrum of the mass operator is discrete, but in the general case
one has also to consider the continuous spectrum (e.g., in the parton model).
 For this reason we
will not specify whether the index enumerating the eigenstates of the
mass operator is discrete or continuous. In the latter case a sum over
$i$ should be understood as an integration.

If $\phi({\bm P}_{\bot},P^+;{\bm k}_1,\sigma_1,...,{\bm k}_N,\sigma_N)
\in {\cal H}_i $
it will be convenient to use the notation
$\phi_i(P;{\bm k}_1,\sigma_1,...,{\bm k}_N,\sigma_N)$ having in mind
that the four-vector
 $P$ has the components $({\bm P}_{\bot},P^+,P_i^-)$, where
$P_i^-=(M_i^2+{\bm P}_{\bot}^2)/2P^+$.
Then, as follows from the comparison of Eqs. (\ref{5}), (\ref{6}) and
(\ref{26}), the action of the representation operators of the
Lorentz group can be written as
\begin{eqnarray}
&&\langle P;S_i,S_{iz}| U(l)|\phi_i\rangle = \sum_{S'_{iz}}
D^{S_i}_{S_{iz}S'_{iz}}[W(l,P_i'/M_i)]\langle P';S_i,S'_{iz}
|\phi_i \rangle
\label{30}
\end{eqnarray}
where $P'=L(l)^{-1}P$.

\section{Current operators in the
front form of dynamics}
\label{S5}

  The translational
covariance of the current operator implies that Eq. (\ref{31})
is satisfied and we can consider this expression as {\it the definition}
of $J^{\mu}(x)$ in terms of $J^{\mu}(0)$. Adopting this definition  it is
easy to show that Poincar\'e covariance of $J^{\mu}(x)$
takes place if i) $J^{\mu}(0)$ satisfies the Lorentz covariance
condition (\ref{33})  and ii) the Poincar\'e group generators satisfy
the condition
(\ref{02}). Therefore the problem of constructing $J^{\mu}(x)$
can be reduced to that of constructing an operator $J^{\mu}(0)$
which satisfies the condition (\ref{33}) and therefore Eq. (\ref{34}).

\begin{sloppypar}
As follows from Eq. (\ref{31}), the continuity equation
$\partial J^{\mu}(x)/\partial x^{\mu} =0$ in terms of
$J^{\mu}(0)$ reads
\begin{equation}
[P_{\mu},J^{\mu}(0)]=0
\label{35}
\end{equation}
\end{sloppypar}

 Eqs. (\ref{34}) and (\ref{35}) show that in the general case
the operator $J^{\mu}(0)$ cannot be chosen the same as for free
particles. Indeed,
the free operator $J^{\mu}_{free}(0)$ obviously satisfies the conditions
(\ref{34}) and (\ref{35}) when the corresponding representation
generators of the Poincar\'e group are free, but in general does not
properly commute with $M^{\mu\nu}$ and $P^{\mu}$, when these operators
are interaction dependent.

 In the front form the set $M^{\mu\nu}$
contains both free and interaction dependent operators and therefore  the
problem of constructing the operator $J^{\rho}(0)$ satisfying
Eq. (\ref{34}) appears to be more complicated than in the point form,
where only $P^{\mu}$ contains the interaction \cite{lev}.
Osborn \cite{Osb} used this
equation to express the operators $J^j(0)$ and
$J^-(0)$ in terms of $J^+(0)$ and then he obtained
a restriction on the kernel of the latter operator. This restriction,
which was called the angular condition (as well as the condition (\ref{1})),
involves triple commutators and therefore it is difficult to  solve.
As shown in Ref. \cite{KoSt}, the Osborn angular condition is automatically
satisfied to first order in $Q$, but  it
is not satisfied to second order if $J^+(0)$ is free. In what follows we
will consider the constraints on the whole set of  components of the
current, adopting a spectral decomposition of the current operator.
Such a procedure allows one   to overcome the difficulties related to
the presence of the interaction in $M^{\mu\nu}$, since the dependence
upon the interacting mass operator becomes a dependence upon  its
eigenvalues.

 If extended Poincar\'e covariance is required (as it is the case for the
em current operator), the
operator $J^{\mu}(0)$ should also properly commute with the operators
$U_{\cal P}$ and $U_{\cal T}$ which, as explained in the preceding section,
are interaction dependent. However, as explained in that
section, the usual Poincar\'e covariance and the proper
commutation relations with $U_y$ and $U_{\theta}$ guarantee that
extended Poincar\'e covariance takes place. Therefore the operator
$J^{\mu}(0)$ should satisfy the conditions
\begin{equation}
U_y J^{\mu}(0) U_y^{-1}=(\Lambda_y)^{\mu}_{\nu}J^{\nu}(0),
\label{36}
\end{equation}
\begin{equation}
U_{\theta}J^{\mu}(0)U_{\theta}^{-1}=J^{\mu}(0)
\label{37} \end{equation}
where the only nonzero components of the matrix $\Lambda_y$ are
$$(\Lambda_y)^0_0=(\Lambda_y)^1_1=-(\Lambda_y)^2_2=(\Lambda_y)^3_3=1.$$

\begin{sloppypar}
 The action of the operator $J^{\mu}(0)$ can be written in the form
\begin{eqnarray}
&&\langle {\bm P}_{\bot},P^+|J^{\mu}(0)|\varphi\rangle =\nonumber\\
&&\int\nolimits J^{\mu}({\bm P}_{\bot},P^+;{\bm P}_{\bot}',P^{'+})
\langle {\bm P}_{\bot}',P^{'+}|\varphi\rangle d\rho({\bm
P}_{\bot}',P^{'+}) \label{38}
\end{eqnarray}
where the kernel
$J^{\mu}({\bm P}_{\bot},P^+;{\bm P}_{\bot}',P^{'+})$ is an
operator in ${\cal H}_{int}$ at any fixed value of its arguments
and the projection $\langle {\bm P}_{\bot},P^+|\varphi\rangle$ is a state
belonging to ${\cal H}_{int}$.
As follows from this expression,
the operator $J^{\mu}(0)$ will be selfadjoint if
\begin{equation}
J^{\mu}({\bm P}_{\bot},P^+;{\bm P}_{\bot}',P^{'+})^*=
J^{\mu}({\bm P}_{\bot}',P^{'+};{\bm P}_{\bot},P^+)
\label{38a}
\end{equation}
where * means
the Hermitian conjugation in ${\cal H}_{int}$ (in the general case the
property of an operator to be selfadjoint is stronger than to be
Hermitian, but we shall not discuss this question).
\end{sloppypar}

\begin{sloppypar}
As mentioned above, the key property that allows one to generalize to
the front form (and also to the instant form) the approach of
\cite{lev} is the following spectral decomposition of the current
operator, viz.
\begin{equation}
J^{\mu}(0)=\sum_{ij}\Pi_i J^{\mu}(0)\Pi_j=\sum_{ij}
J^{\mu}(M_i,M_j)
\label{39}
\end{equation}
where
\begin{equation}
J^{\mu}(M_i,M_j)\equiv  \Pi_i J^{\mu}(0) \Pi_j
\label{40}
\end{equation}
is the part of $J^{\mu}(0)$ describing the transition from
${\cal H}_j$ to ${\cal H}_i$  (for the sake of brevity we do not write the
arguments $S_j$ and $S_i$). If $ \langle {\bm
P}_{\bot},P^+|\varphi_j\rangle\in {\cal H}_j$,
then we can reexpress Eq. (\ref{38}) in the form
\begin{eqnarray}
&&\langle {\bm P}_{\bot},P^{+}|J^{\mu}(M_i,M_j)|\varphi_j\rangle=
\int\nolimits J^{\mu}(P_i;P_j')\nonumber\\
&&\langle {\bm P}_{\bot}',P^{'+}|\varphi_j\rangle
d\rho({\bm P}_{\bot}',P^{'+})
\label{41}
\end{eqnarray}
where
\begin{eqnarray}
&&J^{\mu}(P_i;P_j')\equiv\langle {\bm P}_{\bot},P^+|\Pi_iJ^{\mu}(0)\Pi_j|
{\bm P}_{\bot}',P^{'+}\rangle =\nonumber\\
 && J^{\mu}({\bm
P}_{\bot},P^+,M_i; {\bm P}_{\bot}',P^{'+},M_j)=\Pi_i J^{\mu}
({\bm P}_{\bot},P^+;{\bm P}_{\bot}',P^{'+}) \Pi_j
\label{42}
\end{eqnarray}
Therefore the operator $J^{\mu}(0)$ is fully defined by
the set of the operators $J^{\mu}(P_i,P_j')$, with definite values of
the masses. At any fixed
values of $({\bm P}_{\bot}',P^{'+})$ and $({\bm P}_{\bot},P^+)$
these operators act from ${\cal H}_{j,int}$ to ${\cal H}_{i,int}$,
where ${\cal H}_{i,int}=\Pi_i {\cal H}_{int}$. Since the spaces
${\cal H}_i$ are invariant under the action of the representation
operators of the extended Poincar\'e group, the restrictions imposed
on the operator $J^{\mu}(0)$ by extended Poincar\'e covariance and
current conservation (see Eq. (\ref{35})) can be formulated in
terms of $J^{\mu}(P_i;P_j')$. As follows from Eqs. (\ref{38a}),
(\ref{40}),
(\ref{41}) and (\ref{42}), the operator $J^{\mu}(0)$ will be selfadjoint if
\begin{equation}
J^{\mu}(P_i,P_j')^*=J^{\mu}(P_j',P_i)
\label{43}
\end{equation}
\end{sloppypar}

\section{Extended Lorentz covariance of the current operator}
\label{S6}

\begin{sloppypar}
 As explained in the preceding section, the problem of constructing a
Poincar\'e covariant operator $J^{\mu}(x)$ can be reduced
to that of constructing a Lorentz covariant operator
$J^{\mu}(0)$. This operator is fully defined by the set of
operators $J^{\mu}({\bm P}_{\bot},P^+;{\bm P}_{\bot}',P^{'+})$ (which
in turn are defined by the set of  operators $J^{\mu}(P_i;P_j')$)
{\em acting  through the internal variables}.

First of all, let us consider the covariance with respect to continuous
Lorentz transformations. From  relativistic invariance of $d\rho({\bm
P}_{\bot}',P^{'+})$ and from Eqs. (\ref{33}), (\ref{Wigner}), (\ref{30}),
(\ref{40})
and (\ref{41}), the operator $J^{\mu}(0)$ will be Lorentz covariant if
$J^{\mu}(P_i;P_j')$ fulfills the following relation
\begin{eqnarray}
&&L(l)^{\mu}_{\nu}J^{\nu}(L(l)^{-1}P_i,L(l)^{-1}P_j')=
D^{S_i}[W(l,L(l)^{-1}\frac{P_i}{M_i})]^{-1}\cdot\nonumber\\
&&J^{\mu}(P_i,P_j')D^{S_j}[W(l,L(l)^{-1}\frac{P_j'}{M_j})]
\label{44}
\end{eqnarray}
for (almost) all values of $({\bm P}_{\bot}',P^{'+})$ and
$({\bm P}_{\bot},P^+)$. If $l$ is a front-form boost, i.e. $l=b\in B$
(see Sect. \ref{S2}), the Wigner rotation becomes the identity and then, as
follows from Eq.  (\ref{44}), one has
 \begin{equation}
L(b)^{\mu}_{\nu}J^{\nu}(L(b)^{-1}P_i,
L(b)^{-1}P_j')=J^{\mu}(P_i,P_j')
\label{45}
\end{equation}
\end{sloppypar}

\begin{sloppypar}
In order to investigate in detail  the constraints imposed on
$J^{\mu}(P_i;P_j')$ by
 Lorentz covariance, it is convenient to consider the current in a general
 Breit frame, and then the current in a particular
Breit frame where the three-momentum is directed along the $z$ axis.
In the latter
frame, as will be clear in what follows, one can take
advantage of the rotational symmetry, differently from the case where the
frame $q^+=0$ is chosen. It  is worth noting that the definition of
a Breit frame is possible because the masses are well defined in both
the initial and final states. (In the point form, cf. \cite{lev}, the
construction of the covariant current was carried out in the
equal-velocity frame, and therefore it was not necessary to fix the
masses). The Breit frame is defined as the reference frame where the
initial and final momenta are
\begin{equation}
K_i=B(H_{ij})^{-1}P_i, \qquad K_j'=B(H_{ij})^{-1}P_j'
\label{46}
\end{equation}
In Eq. (\ref{46}) $H_{ij} \equiv (P_i+P_j')/|P_i+P_j'|$ and
$B(H_{ij})$  denotes the Lorentz transformation $L[\beta(H_{ij})]$. The
four-vectors $K_i$ and $K_j'$ in Eq. (\ref{46}) are such that
\begin{equation}
K_i^2=M_i^2, \quad K_j^{'2}=M_j^2, \quad {\bm K}_i+{\bm K}_j'=0
\label{47}
\end{equation}
Therefore the four-vectors $K_i$ and
$K_j'$ are fully determined by one three-dimensional vector
${\bm K}_{ij}\equiv {\bm K}_i$.
The relations (\ref{47}) can also be directly obtained from Eqs.
(\ref{4}) and
(\ref{46}), since, as follows from these expressions (compare with
Eq. (\ref{16})) one has
\begin{equation}
K_i^+=\frac{P_i^+}{\sqrt{2}H_{ij}^+},\quad {\bm K}_{i\bot}=
{\bm P}_{i\bot}-\sqrt{2}K_i^+{\bm H}_{ij\bot}
\label{48}
\end{equation}
and $K_j'$ is given by the same expressions with $P_i$ replaced by
$P_j'$.
\end{sloppypar}

\begin{sloppypar}
 As follows from Eqs. (\ref{45}) and (\ref{46}),
\begin{equation}
J^{\mu}(P_i,P_j')=B(H_{ij})^{\mu}_{\nu}j^{\nu}({\bm K}_{ij};M_i,M_j)
\label{49}
\end{equation}
where we use
$j^{\nu}({\bm K}_{ij};M_i,M_j)$ to denote $J^{\mu}(K_i,K_j')$, i.e. the
current in the Breit frame.  From Eqs. (\ref{43}) and
(\ref{49}), the condition for the
Hermiticity of the operator $J^{\mu}(P_i,P_j')$ will be satisfied
if and only if
\begin{equation}
j^{\nu}({\bm K}_{ij};M_i,M_j)^*=j^{\nu}(-{\bm K}_{ij};M_j,M_i)
\label{50}
\end{equation}
\end{sloppypar}

\begin{sloppypar}
 Since the operator $J^{\mu}(P_i,P_j')$ can be expressed in terms of
$j^{\nu}({\bm K}_{ij};M_i,M_j)$, we will look for   the properties
of $j^{\nu}({\bm K}_{ij};M_i,M_j)$ such that the
operator $J^{\mu}(P_i,P_j')$ defined by Eq. (\ref{49})
satisfies Eq. (\ref{44}). This latter becomes in terms of
$j^{\nu}({\bm K}_{ij};M_i,M_j)$
\begin{eqnarray}
&&j^{\mu}(L[W^{-1}(l,L(l)^{-1}H_{ij})]{\bm K}_{ij};M_i,M_j)=
L[W^{-1}(l,L(l)^{-1}H_{ij})]^{\mu}_{\nu}\cdot\nonumber\\
&&D^{S_i}[W^{-1}(l,L(l)^{-1}\frac{P_i}{M_i})]
j^{\nu}({\bm
K}_{ij};M_i,M_j)
D^{S_j}[W^{-1}(l,L(l)^{-1}\frac{P_j'}{M_j})]^{-1}
\label{51'}
\end{eqnarray}

Let us define $u\in SU(2)$ as follows
\begin{eqnarray}
&& u=W^{-1}(l,L(l)^{-1}H_{ij})
\label{51}
\end{eqnarray}
As shown in Appendix A, from Eq. (\ref{51}) one has
\begin{equation}
 W^{-1}(l,L(l)^{-1}{P_i \over M_i})= W(u,{K_i \over M_i})~,
 \quad  W^{-1}(l,L(l)^{-1}{P'_j \over M_j})= W(u,{K'_j \over M_j})
\label{52}
\end{equation}
Therefore Eq. (\ref{51'}) becomes
\begin{eqnarray}
&&j^{\mu}(L(u){\bm K}_{ij};M_i,M_j)=L(u)^{\mu}_{\nu}
D^{S_i}[W(u,\frac{K_i}{M_i})]\cdot\nonumber\\
&&j^{\nu}({\bm K}_{ij};M_i,M_j)D^{S_j}[W(u,\frac{K_j'}{M_j})]^{-1}
\label{53}
\end{eqnarray}

It is clear that if Eq. (\ref{53}) is satisfied for any
$u\in SU(2)$, then  Eq. (\ref{51'}) is satisfied too. Therefore we can
investigate Eq. (\ref{53}) only.

 As follows from Eqs. (\ref{Wigner}), (\ref{12}) and (\ref{13}), Eq.
(\ref{53}) can be written in the form
\begin{eqnarray}
&&j^{\mu}(L(u){\bm K}_{ij};M_i,M_j)=L(u)^{\mu}_{\nu}
D^{S_i}[v(L(u)\frac{K_i}{M_i})^{-1}uv(\frac{K_i}{M_i})]\cdot\nonumber\\
&&j^{\nu}({\bm K}_{ij};M_i,M_j)D^{S_j}[v(\frac{K_j'}{M_j})^{-1}u^{-1}
v(L(u)\frac{K_j'}{M_j})]
\label{54}
\end{eqnarray}
where the Melosh rotations $v$ appear instead of the boosts $\beta$
(see Eq.(\ref{13})). This replacement will be useful in the following.
Note that when ${\bm K}_{ij}=0$ Eq. (\ref{54}) becomes
\begin{equation}
j^{\mu}(\bm{0};M_i,M_j)=L(u)^{\mu}_{\nu}D^{S_i}(u)
j^{\nu}(\bm{0};M_i,M_j)D^{S_j}(u)^{-1}
\label{55}
\end{equation}

\begin{sloppypar}
 We conclude that the operator $J^{\mu}(P_i,P_j')$
will satisfy the Lorentz covariance condition (\ref{44})
if the operator $j^{\nu}({\bm K}_{ij};M_i,M_j)$ satisfies the rotational
covariance condition (\ref{54}) for any $u$. In the point form, the equation
analogous  to  Eq. (\ref{54}) does not contain Melosh matrices,
but  the relation between
$J^{\mu}$ and $j^{\nu}$ is more
complicated than  Eq. (\ref{49})  \cite{lev}.
\end{sloppypar}

 Equation (\ref{54}) can be used for  expressing
$j^{\nu}({\bm K};M_i,M_j)$,  corresponding to an arbitrary
three-momentum ${\bm K}$, in terms of   auxiliary operators
$j^{\mu}(K{\bm e}_z;M_i,M_j)$, which are constrained only by covariance
relative  to rotations around the $z$ axis, $u_z$, and not
 by the full SU(2) covariance. The  auxiliary operators
 $j^{\mu}(K{\bm e}_z;M_i,M_j)$ represent the current in a Breit frame where
 ${\bm K}$ is along the $z$ axis. If
${\bm K}=K {\bm e}_z$, where $K=|{\bm K}|$ and ${\bm
e}_z$ is the unit vector along the positive direction of the $z$ axis, then
from (\ref{54}) one has for rotations around the $z$ axis

\begin{equation}
j^{\mu}(K{\bm e}_z;M_i,M_j)=L(u_z)^{\mu}_{\nu}
exp(-\imath \varphi S^z_i)j^{\nu}(K{\bm e}_z;M_i,M_j)exp(\imath
\varphi S^z_j)
\label{56}
\end{equation}
where $exp(-\imath \varphi S^z_{i(j)})=D^{S_{i(j)}}(u_z)$ and the
relation $v(g) =1$ for ${\bm g}_{\perp}=0 $ has been used (see  Eq.
(\ref{13})).
From Eq. (\ref{56}), it is clear that both $j^+$ and $j^-$ must be
diagonal with respect
to the third component of the spin, while ${\bm j}_{\perp}$ should
properly transform with respect to the rotations $u_z$.

\begin{sloppypar}
In order to demonstrate that the auxiliary operators are constrained only by
covariance relative to rotations around the $z$ axis,
let $r({\bm K})\in SU(2)$ be such that
$L[r({\bm K})]K {\bm e}_z={\bm K}$. In particular, if $\varphi$
and $\theta$ are the polar angles characterizing the vector ${\bm K}$,
then $r({\bm
K})$ can be chosen in the form
\begin{equation}
r({\bm K})=exp(-\frac{\imath}{2}\varphi \tau_3)
exp(-\frac{\imath}{2}\theta \tau_2).
\label{58}
\end{equation}

Then,
from Eq. (\ref{54}), replacing ${\bm K}_{ij}$ with $K_{ij}{\bm e}_z$
and $u$ with $r({\bm K}_{ij})$ one has
\begin{eqnarray}
&&j^{\mu}({\bm K}_{ij};M_i,M_j)=L[r({\bm K}_{ij})]^{\mu}_{\nu}
D^{S_i}[v(\frac{K_i}{M_i})^{-1}r({\bm K}_{ij})]\cdot\nonumber\\
&&j^{\nu}(K_{ij}{\bm e}_z;M_i,M_j)D^{S_j}[r({\bm K}_{ij})^{-1}
v(\frac{K_j'}{M_j})]
\label{57}
\end{eqnarray}
where the property of the Melosh rotations that $v(g)=1$ for
${\bm g}_{\perp}=0 $ has
been used once more.
Now we consider this expression as the definition of
$j^{\mu}({\bm K}_{ij};M_i,M_j)$ in terms of
$j^{\mu}(K_{ij}{\bm e}_z;M_i,M_j)$.
Note that this definition is meaningful only if $K_{ij}\neq 0$.
Using Eq. (\ref{57}), it is easy to show that Eq. (\ref{54}) in
terms of $j^{\mu}(K_{ij}{\bm e}_z;M_i,M_j)$ becomes
\begin{eqnarray}
&&j^{\mu}(K_{ij}{\bm e}_z;M_i,M_j)=L[r(L(u)
{\bm K}_{ij})^{-1}ur({\bm K}_{ij})]^{\mu}_{\nu}\cdot\nonumber\\
&&D^{S_i}[r(L(u){\bm K}_{ij})^{-1}ur({\bm K}_{ij})]
j^{\nu}(K_{ij}{\bm e}_z;M_i,M_j)\cdot\nonumber\\
&&D^{S_j} [r(L(u){\bm K}_{ij})^{-1}ur({\bm K}_{ij})]^{-1}
\label{57'}
\end{eqnarray}
From the properties of the products of rotation operators
(see, e.g., \cite{VMK}), one obtains $r(L(u){\bm K}_{ij})^{-1}
ur({\bm K}_{ij}) = u_z$, where $u_z$ is a well-defined rotation
around the $z$ axis. Therefore,  if $j^{\mu}(K{\bm e}_z;M_i,M_j)$
satisfies Eq. (\ref{56}) for any $u_z$, then  $j^{\mu}({\bm K};M_i,M_j)$,
defined by Eq. (\ref{57}),
fulfills Eq. (\ref{54}). This means that, in order to fulfill
Poincar\'e covariance, $ J^{\mu}(P_i,P_j')$ can be expressed in
terms of the auxiliary operators $j^{\mu}(K{\bm e}_z;M_i,M_j)$ constrained
only by rotations around the $z$ axis.
\end{sloppypar}

\begin{sloppypar}

 The property of Hermiticity  (cf. Eq. (\ref{50})) for
$j^{\mu}(\bm{0};M_i,M_j)$ reads
\begin{eqnarray}
&&j^{\mu}({\bm 0};M_i,M_j)^*=j^{\mu}({\bm 0};M_j,M_i)
\label{60'}
\end{eqnarray}
while for $|{\bm K}|\neq 0$,  from Eq. (\ref{57}), the property of
Hermiticity  for
$j^{\mu}(K\bm{e}_z;M_i,M_j)$ becomes
\begin{eqnarray}
&&j^{\mu}(K{\bm e}_z;M_i,M_j)^*= L(r({\bm K})^{-1}
r(-{\bm K}))^{\mu}_{\nu} D^{S_j}[r({\bm K})^{-1}r(-{\bm K})] \nonumber \\
&&j^{\nu}(K{\bm e}_z;M_j,M_i)D^{S_i}[r({\bm K})^{-1}
r(-{\bm K})]^{-1}= \nonumber \\
&&L[r_x(-\pi)]^{\mu}_{\nu} D^{S_j}[r_x(-\pi)]
j^{\nu}(K{\bm e}_z;M_j,M_i) D^{S_i}[r_x(-\pi)])^{-1},
\label{60}
\end{eqnarray}
 since from Eq. (\ref{58}) one has
\begin{equation}
r({\bm K})^{-1}r(-{\bm K})=exp({\imath}{\pi \over 2}\tau_1)=\imath
\tau_1\equiv r_x(-\pi), \label{59}
\end{equation}
where $r_x(-\pi) \in SU(2)$ yields the rotation by $-\pi$ around the
$x$ axis.
It is worth noting that Eq. (\ref{60}) represents a non trivial
constraint  when $M_i=M_j$ (i.e., for  elastic form factors), because
in this case the rhs and the
lhs contain the same operator. In the inelastic case, one could
construct the current defining the operator $j^{\mu}(K{\bm e}_z;M_j,M_i)$ for
$M_i < M_j$ and use Eq. (\ref{60}) in order
to extend the definition of the current also for $M_i > M_j$. 
Another possible choice is to define the current for any value of the masses; 
in this case    Eq.(\ref{60}) becomes  a non trivial
constraint also in  inelastic processes. We will call {\em Hermiticity
condition} the relation given by Eq. (\ref{60}).
\end{sloppypar}

\begin{sloppypar}
 Let us now summarize the above results. The current
operator $J^{\mu}(0)$ satisfies Lorentz covariance if the
operator $J^{\mu}(P_i,P_j')$ satisfies the condition (\ref{44}).
This condition is satisfied if i) $J^{\mu}(P_i,P_j')$ is defined
by Eq. (\ref{49}), ii)  $j^{\mu}({\bm K};M_i,M_j)$ is defined by Eq.
(\ref{57}), if ${\bm K}\neq 0$, and satisfies Eq. (\ref{55}) if
${\bm K}=0$, and iii) the auxiliary operators
$j^{\mu}(K{\bm e}_z;M_i,M_j)$ satisfy Eq.
(\ref{56}). In addition, Eqs. (\ref{60'}) and (\ref{60})  guarantee that
the operator $ J^{\mu}(0)$ constructed in such a way is Hermitian.
\end{sloppypar}

 Of course it is possible to choose another set of minimally
constrained operators by choosing ${\bm K}$ along any other axis.
However the choice of the reference frame where ${\bm K}= K{\bm e}_z$ is
the most convenient, since the rotations around the $z$ axis are
interaction independent
and furthermore  there are no Melosh matrices in Eqs. (\ref{56}) and
(\ref{60}) (this follows from the fact that $v(g)=1$ if ${\bm g}$ is
directed along the $z$ axis). It is worth noting that the continuous
Lorentz transformations constrain the current
$j^{\mu}(K{\bm e}_z;M_i,M_j)$ for a non-interacting
system in the same way as in the
interacting case, namely Eq. (\ref{56}) holds for both non-interacting and
interacting systems, since rotations around the $z$ axis are
interaction free.

 For the em current,  also ${\cal P}$ and ${\cal T}$ covariance is
required, i.e. extended Lorentz covariance is needed.
 As explained in Sects. \ref{S4} and \ref{S5}, the current operator
satisfies ${\cal P}$ covariance if it satisfies Poincar\'e covariance
and Eq. (\ref{36}). As follows from Eqs. (\ref{22}), (\ref{23}),
(\ref{38}-\ref{42}), (\ref{48}) and (\ref{49}), the condition
(\ref{36}) is satisfied if and only if
\begin{equation}
(\Lambda_y)^{\mu}_{\nu}j^{\nu}( {\bm {\tilde K}},M_i,M_j)=
U_{y,int}j^{\mu}({\bm K},M_i,M_j)U_{y,int}^{-1}
\label{71}
\end{equation}
where $ {\bm {\tilde K}} \equiv (K_x,-K_y,K_z)$

\begin{sloppypar}
As follows from Eq. (\ref{57}), this condition in
terms of $j^{\mu}(K{\bm e}_z;M_i,M_j)$ reads
\begin{equation}
(\Lambda_y)^{\mu}_{\nu}j^{\nu}(K{\bm e}_z,M_i,M_j)=
U_{y,int}j^{\mu}(K{\bm e}_z;M_i,M_j)U_{y,int}^{-1}
\label{72}
\end{equation}
where the properties $\Lambda_y~ L[r(\bm{\tilde{K}})]
\Lambda_y^{-1} = L[r(\bm{{K}})]$ and
$U_{y,int}~D^{S_i}[v(K_i/M_i)^{-1}r({\bm K})]~U_{y,int}^{-1}=
D^{S_i}[v(\tilde{K}_i/M_i)^{-1}r({\bm {\tilde{K}}})]$ have been used.
\end{sloppypar}

 Analogously, as follows from Eqs. (\ref{25}), (\ref{38}-\ref{42}),
and (\ref{49}), the condition (\ref{37}), which guarantees $\theta$
covariance, is satisfied if and only if
\begin{eqnarray}
& & j^{\mu}(K{\bm e}_z;M_i,M_j)=U_{\theta}{j}^{\mu}(K{\bm e}_z;M_i,M_j)
U_{\theta}^{-1}
\label{73}
\end{eqnarray}

\begin{sloppypar}
The constraints imposed on the  current for an interacting system by
extended Lorentz covariance can be fulfilled by a current composed in
our Breit
frame by the sum of only one-body  currents (e.g. $\sum_{i=1}^{N}
j_{free,i}^{\mu}$, where N is the number of  constituents in the
interacting system), while some additional care must be adopted for the
Hermiticity (cf. Eq. (\ref{60}) and Sects. \ref{S9} - \ref{S10}). The
extended Lorentz covariance is clearly satisfied by the one-body
free current, since the constraints   are the same for a
non-interacting and an interacting system (cf. Sect. \ref{S4}). It is
worth noting that the same analysis  performed in our Breit frame
can be carried out in any other reference frame obtained by a boost
along the $z$ axis, since the symmetry around the $z$ axis is preserved.

 If  the cluster separability is important, i.e.  $A\neq 1$, one can
adopt $A j_{free}^{\mu}A^{-1}$ (as discussed in
Sects. \ref{S9} - \ref{S10} and in \cite{lev,LPS}) in order to construct a
current which fulfills
extended Lorentz covariance and  Hermiticity.

\end{sloppypar}

\section{Current conservation and charge operator}
\label{S7}

  The results of the
preceding section give the full solution of the problem of constructing
the current in the front form as far as  Poincar\'e covariance and
Hermiticity are concerned. However the em current operator
(differently from the weak one) should also satisfy  current
conservation and a proper normalization condition in terms of
 the electric charge of the system.

 As follows from Eqs. (\ref{35})and (\ref{39}-\ref{42}), the continuity
 equation will be satisfied if
\begin{equation}
(P_i-P_j')_{\mu}J^{\mu}(P_i,P_j')=0,
\label{61}
\end{equation}
In the Breit frame Eq. (\ref{61}) (cf. Eq. (\ref{49})) becomes
\begin{eqnarray}
&&(K_i^- - {K'}_j^{-})j^+(\bm {K};M_i,M_j)+(K_i^+ -
{K'}_j^{+})j^-(\bm{ K};M_i,M_j)- \nonumber\\
&&2\bm{K}_{\perp} \cdot \bm{j}_{\perp}(\bm {K};M_i,M_j)=0
\label{62}
\end{eqnarray}
where $K^+_{i}=(\sqrt{M^2_{i}+|\bm{K}|^2}+K_z)/\sqrt{2}$ and
${K'}^{+}_{j}=(\sqrt{M^2_{j}+|\bm{K}|^2}-K_z)/\sqrt{2}$, while
$K^-_{i}= (M^2_{i}+|\bm{K}_{\perp}|^2)/(2 K^+_{i})$
and ${K'}^{-}_{j}= (M^2_{j}+|\bm{K}_{\perp}|^2)/(2 {K'}^{+}_{j})$.

If $\bm {K}_{\perp}= 0$ and $ {K}_{z}\neq 0$, then Eq. (\ref{62}) yields
\begin{equation}
({M^2_{i} \over 2 K^+_{i}} - {M^2_{j} \over 2 {K'}^+_{j}})j^+
(K \bm {e}_z;M_i,M_j)+(K_i^+ - {K'}_j^+)j^-(K\bm {e}_z;M_i,M_j)=0
\label{63}
\end{equation}
while, if $\bm{K}= 0$ and $M_i \neq M_j$, then from Eq. (\ref{63}) we have
\begin{equation}
j^-(\bm{0};M_i,M_j)=-j^+(\bm{0};M_i,M_j).
\label{64}
\end{equation}
By taking the derivatives of Eq. (\ref{62}) at $\bm{K}= 0$, one has
\begin{eqnarray}
&&{\bm j}_{\perp}(\bm{0};M_i,M_j)=\frac{1}{2\sqrt{2}} (M_i-M_j)
\frac{\partial}{\partial {\bm K}_{\perp}} \left [j^+(\bm{0};M_i,M_j)+
j^-(\bm{0};M_i,M_j)\right ], \nonumber \\
&&\left [j^+(\bm{0};M_i,M_j)-j^-(\bm{0};M_i,M_j)\right ] = \nonumber \\
&&\frac{1}{2}(M_i-M_j)
\frac{\partial}{\partial { K}_{z}}
\left [j^+(\bm{0};M_i,M_j)+j^-(\bm{0};M_i,M_j)\right ]
\label{65}
\end{eqnarray}
In particular if $M_i=M_j$ one has from Eq. (\ref{65})
\begin{eqnarray}
&&{\bm j}_{\perp}(\bm{0};M_i,M_i)=0 \nonumber \\
&&j^+(\bm{0};M_i,M_i)=j^-(\bm{0};M_i,M_i)
\label{66}
\end{eqnarray}
Note that the  signs in Eq. (\ref{64}) (inelastic case)
and in the second line of  (\ref{66}) (elastic case) differ each other.
\begin{sloppypar}
As follows from Eq. (\ref{63}), if $K\neq 0$
then only $j^+(K\bm{e}_z;M_i,M_j)$ and $j^-(K\bm{e}_z;M_i,M_j)$ are
 constrained by
the continuity equation, viz.
\begin{equation}
j^-(K\bm{e}_z;M_i,M_j)=-{\left [ M^2_{i} /( 2 K^+_{i})  - M^2_{j}
 / (2 {K'}^+_{j}) \right ] \over (K_i^+ - {K'}_j^+)} j^+(K\bm{e}_z;M_i,M_j).
\label{67}
\end{equation}
while $\bm{j}_{\perp}(K\bm{e}_z;M_i,M_j)$
remains unconstrained.
If we choose in our Breit frame
$j^+(K\bm{e}_z;M_i,M_j)$ and $\bm{j}_{\perp}(K\bm{e}_z;M_i,M_j)$
free, then, as follows from Eq. (\ref{67}),
$j^-(K\bm{e}_z;M_i,M_j)$ must be interaction dependent, because of   current
conservation. However, in the actual calculations of any elastic and
inelastic form factor only three components of the current are
necessary (cf. Sects. \ref{S9} and \ref{S10}), and these components can be
chosen  free.
\end{sloppypar}

 Let us now consider the charge operator. In front-form dynamics
it is defined as
\begin{equation}
Q=\int\nolimits J^+(x) \delta(x^+)d^4x
\label{68}
\end{equation}
Therefore,  from Eqs. (\ref{31}) and (\ref{38}), one has
\begin{eqnarray}
&&\langle {\bm P}_{\bot},P^+|Q|\varphi\rangle  =
\int\nolimits \int\nolimits \int\nolimits
J^+({\bm P}_{\bot},P^+;
{\bm P}_{\bot}',P^{'+})\cdot\nonumber\\
&& exp~\imath [(P^+-P^{'+})x^- -({\bm P}_{\bot}-{\bm P}_{\bot}') \cdot
{\bm x}_{\bot}]\langle {\bm P}'_{\bot},P^{'+}|
\varphi\rangle\cdot \nonumber \\
&&d{\bm x}_{\bot}dx^- d\rho({\bm P}_{\bot}',P^{'+})=
\sum_{ij} \frac{1}{2P^+}
J^+(P_i,P_j)\langle {\bm P}_{\bot},P^{+}|\varphi\rangle
\label{69}
\end{eqnarray}
where  Eqs. (\ref{39}) and (\ref{42}) have been used.
 From Eq. (\ref{61}), for $P^{'+}=P^{+}$ and ${\bm P}_{\bot}'=
{\bm P}_{\bot}$, one has $(P_i^- -P_j^-)J^+(P_i,P_j)=0$, and then
$J^+(P_i,P_j)=0$, if $M_i\neq M_j$.
Therefore only the terms with $i=j$ contribute to the sum in
Eq. (\ref{69}). Since  from Eqs. (\ref{46}) and (\ref{48}),
$B^+_+(H_{ii})=\sqrt{2}H_{ii}^+$,
and all the other components of $B^+_{\nu}(H_{ii})$ are equal to zero,
using Eq. (\ref{49}), $J^+(P_i,P_i)$ can be expressed in terms of
$j^{+}(\bm{K}_{ij};M_i,M_i)$, where in our case $\bm{K}_{ij}=0$.
Therefore,  Eq.
(\ref{69}) becomes
\begin{equation}
\langle {\bm P}_{\bot},P^+|Q|\varphi\rangle =\sum_{i}\frac{1}
{\sqrt{2}M_i} j^+(\bm{0};M_i,M_i)
\langle {\bm P}_{\bot},P^+|\varphi\rangle
\label{70}
\end{equation}

 We conclude that for each subspace ${\cal H}_i$ one must have (cf.
Eq. (\ref{66}))
\begin{equation}
j^+(\bm{0};M_i,M_i)=j^-(\bm{0};M_i,M_i)={\sqrt{2}}eM_i \Pi_i
\label{70a}
\end{equation}
where $e$ is the total electric charge of the
system under consideration. This normalization condition is trivially
fulfilled by $\sum_{i=1}^N j_{free,i}^{+}$.

All the above results show that the operator $J^{\mu}(0)$ satisfies
i) Lorentz covariance, ii) ${\cal P}$ and
${\cal T}$ covariance, iii) Hermiticity, iv) continuity equation and v)
charge conservation, if the operator $j^{\mu}(K \bm{e}_z;M_i,M_j)$
satisfies Eqs. (\ref{55}),
(\ref{56}), (\ref{72}), (\ref{73}), (\ref{60'}), (\ref{60}), (\ref{63}) and
(\ref{70a}).
 However, even if all these
conditions are satisfied, this does not guarantee that the current operator
fulfills  locality and cluster separability
(cf. Sect. \ref{S1} and Ref. \cite{LPS}).

\section{Matrix elements of the current operator}
\label{S8}

 In the scattering theory, one-particle states with
four-momentum $p'$ and  spin projection $\sigma'$ are usually
normalized as
\begin{equation}
\langle p",\sigma"|p',\sigma'\rangle=2(2\pi)^3p^{'+}
\delta^{(2)}({\bm p}_{\bot}"-
{\bm p}_{\bot}')\delta(p^{"+}-p^{'+})\delta_{\sigma"\sigma'}
\label{75}
\end{equation}
where $\delta_{\sigma\sigma'}$ is the Kronecker symbol.

 Since the form of the generators (\ref{26}) is analogous to the
form of the one-particle generators (\ref{6}), the wave function of
the state with  four-momentum $P'$ and
internal wave function $\chi'$ can be written in the form
\begin{eqnarray}
&&\langle {\bm P}_{\bot},P^+;{\bm k}_1,\sigma_1,...,{\bm k}_N,\sigma_N
|P',\chi'\rangle =
2(2\pi)^3P^{'+}\delta^{(2)}({\bm P}_{\bot}-
{\bm P}_{\bot}')\cdot\nonumber\\
&&\delta(P^+-P^{'+})\chi'({\bm k}_1,\sigma_1,...,{\bm k}_N,\sigma_N )
\label{76}
\end{eqnarray}
Indeed, as follows from Eqs. (\ref{19}) and (\ref{18}), in this case
the normalization is analogous to that in Eq. (\ref{75}):
\begin{eqnarray}
&&\langle P",\chi"|P',\chi'\rangle =
2(2\pi)^3P^{'+}\delta^{(2)}({\bm P}_{\bot}"-
{\bm P}_{\bot}')
 \delta(P^{"+}-P^{'+})\,\langle\chi"|\chi'\rangle
\label{77}
\end{eqnarray}
where the scalar product on the right-hand side is understood only
in the space ${\cal H}_{int}$.

 Let us now consider the em or weak transition of the
state with  mass $M_j$, four-momentum $P_j'$ (such that
$P_j^{'2}=M_j^2$) and  internal wave function $\chi_j$ to the
state with  mass $M_i$, four-momentum $P_i$ and  internal
wave function $\chi_i$. Then, as follows from Eqs.  (\ref{31}),
(\ref{38}-\ref{42}) and (\ref{49}),
\begin{eqnarray}
&&\langle P_i,\chi_i|J^{\mu}(x)|P_j',\chi_j\rangle
=exp[\imath (P_i-P_j') x]B(H_{ij})^{\mu}_{\nu}\langle \chi_i|
j^{\nu}({\bm K}_{ij},M_i,M_j)|\chi_j\rangle \nonumber \\
\label{78}
\end{eqnarray}
where the matrix element on the right-hand side must be calculated
only in the space ${\cal H}_{int}$.

From Eq. (\ref{78}), it is clear that a process in an arbitrary frame can be
 investigated in terms of the current in the Breit frame. This observation
has been essentially used in the method of Ref. \cite{PK}.

 In turn, using Eq. (\ref{57}), we can express matrix elements
of the current operator in terms of the matrix elements of the
operator ${ j}^{\mu}(K_{ij} \bm{e}_z;M_i,M_j)$:
\begin{eqnarray}
&&\langle P_i,\chi_i|J^{\mu}(x)|P_j',\chi_j\rangle
=exp[\imath (P_i-P_j') x]\cdot\nonumber\\
&&L[\beta(H_{ij})r({\bm K}_{ij})]^{\mu}_{\nu}\langle \chi_i|
D^{S_i}[v(K_i/M_i)^{-1}r({\bm K}_{ij})]\cdot\nonumber\\
&&j^{\nu}(K_{ij}\bm{e}_z;M_i,M_j)D^{S_j}[r({\bm K}_{ij})^{-1}v(K_j'/M_j)]
|\chi_j\rangle
\label{79}
\end{eqnarray}

As follows from Eqs. (\ref{46}), (\ref{79}) and from the definition of
$\bm{K}_{ij}$, the matrix elements of the current operator in terms of
the matrix elements of the operator $j^{\mu}(K_{ij}\bm{e}_z;M_i,M_j)$
have the  simplest form in the reference frame where
\begin{equation}
{\bm P}_{i\bot}={\bm P}_{j\bot}'=0,\quad P_i^z+
P_j^{'z}=0,\quad P_i^z \ne 0
\label{80}
\end{equation}
Indeed, in this case Eq. (\ref{79}) obviously yields
\begin{equation}
\langle P_i,\chi_i|J^{\mu}(0)|P_j',\chi_j\rangle
=\langle \chi_i|j^{\mu}( \pm K_{ij}\bm{e}_z;M_i,M_j)|\chi_j\rangle
\label{81}
\end{equation}
where $K_{ij}=|P_i^z|$ and  $\pm = sign(P_i^z)$.

\indent  It is useful to investigate  the
matrix elements of the current $j^{\mu}(K\bm{e}_z;M_i,M_j)$ between
different internal states $|\chi_i \rangle$ and $|\chi_j \rangle$, and the
constraints imposed by the covariance for rotations around
the $z$ axis. In the em case, one has to
consider also the constraints imposed by
parity and  time reversal covariance (the current conservation will be
furtherly discussed in Sect. \ref{S10}).
 From Eq. (\ref{56})  one
immediately obtains
\begin{eqnarray}
 && j^{+}(K\bm{e}_z;M_i,M_j) =
exp(-\imath \varphi S_z)j^{+}(K\bm{e}_z;M_i,M_j)
exp(\imath \varphi S_z) \nonumber \\
&& j^{-}(K\bm{e}_z;M_i,M_j) =
exp(-\imath \varphi S_z)j^{-}(K\bm{e}_z;M_i,M_j)
exp(\imath \varphi S_z)
\label{100}
\end{eqnarray}
and therefore only the diagonal matrix elements of
$j^{\pm}(K\bm{e}_z;M_i,M_j)$ are different from zero. As for the
$\perp$ components of the current, one has
\begin{eqnarray}
 && j_x(K\bm{e}_z;M_i,M_j) =
exp(-\imath \varphi S_z) \nonumber \\ && \left [ cos
\varphi~ j_x(K\bm{e}_z;M_i,M_j)+ sin \varphi ~j_y(K\bm{e}_z;M_i,M_j) \right ]
exp(\imath \varphi S_z) \nonumber \\
\nonumber \\
&& j_y(K\bm{e}_z;M_i,M_j) =
exp(-\imath \varphi S_z) \nonumber \\ && \left [ -sin
\varphi~ j_x(K\bm{e}_z;M_i,M_j)+ cos \varphi ~j_y(K\bm{e}_z;M_i,M_j) \right ]
exp(\imath \varphi S_z)
\label{101}
\end{eqnarray}
Let $|\chi_{i(j)} \rangle  = | M_{i(j)} S_{i(j)} S_{iz(jz)} \rangle \in
{\cal H}_{int}$ be an
eigenstate of mass, $M_{i(j)}$, intrinsic angular momentum, $S_{i(j)}$,
and third
component of
angular momentum, $S_{iz(jz)}$.
From Eq. (\ref{101}) it is straightforward to obtain the matrix
elements of $j_y$ from the ones of $j_x$, namely
\begin{eqnarray}
&& \langle  S_{iz} S_i  M_i |j_y(K\bm{e}_z;M_i,M_j)| M_j S_j S_{jz} \rangle =
-exp[-\imath { \pi \over 2} (S_{iz}-S_{jz})] \nonumber \\ && \langle S_{iz}
S_i  M_i |j_x(K\bm{e}_z;M_i,M_j)| M_j S_j S_{jz} \rangle
\label{102}
\end{eqnarray}
Furthermore, after substituting Eq. (\ref{102}) in Eq. (\ref{101}) one finds 
that the  matrix
elements are vanishing unless
$(S_{iz}-S_{jz})^2=1$.

The $U_y$-parity covariance, (Eq. (\ref{72})),  yields
\begin{eqnarray}
 && \langle   S_{iz} S_i  M_i |j^{\pm}(K\bm{e}_z;M_i,M_j)| M_j S_j S_{jz}
 \rangle =  
 \eta_{{\cal{P}}_i}\eta_{{\cal{P}}_j} (-1)^{(S_i-S_j)} \nonumber \\
&&\langle   -S_{iz} S_i  M_i |j^{\pm}(K\bm{e}_z;M_i,M_j)|M_j S_j -S_{jz}
  \rangle, \label{104a} \\
\nonumber
\\
&& \langle  S_{iz} S_i  M_i |j_x(K\bm{e}_z;M_i,M_j)|
M_j S_j S_{jz} \rangle
=~-  \eta_{{\cal{P}}_i}\eta_{{\cal{P}}_j} (-1)^{(S_i-S_j)}\nonumber\\
&&\langle  -S_{iz} S_i
  M_i |j_x(K\bm{e}_z;M_i,M_j)| M_j S_j -S_{jz} \rangle ,\label{104b} \\ 
  \nonumber \\
 && \langle  S_{iz} S_i  M_i |j_y(K\bm{e}_z;M_i,M_j)| M_j S_j S_{jz}
\rangle =
  \eta_{{\cal{P}}_i}\eta_{{\cal{P}}_j}(-1)^{(S_i-S_j)} \nonumber\\
&&\langle  -S_{iz} S_i
  M_i |j_y(K\bm{e}_z;M_i,M_j)| M_j S_j -S_{jz} \rangle \label{104c} 
\end{eqnarray}
and therefore  only the matrix elements with $S_{iz} \geq 0 $ are
independent.

 Finally the $U_{\theta}$-parity ('time reversal') covariance,
(Eq. (\ref{73})),  gives
 \begin{eqnarray}
 && \langle  S_{iz} S_i  M_i |j^{\mu}(K\bm{e}_z;M_i,M_j)| M_j S_j S_{jz}
 \rangle = \eta_{\theta_i}\eta_{\theta_j}
  (-1)^{(S_i+S_j-S_{iz}-S_{jz})} \nonumber \\
&&{\overline {\langle  -S_{iz} S_i  M_i
|j^{\mu}(K\bm{e}_z;M_i,M_j)|M_j S_j -S_{jz}  \rangle}}
\label{105}
 \end{eqnarray}
 As is well known,  combining parity, Eqs. (\ref{104a})-(\ref{104b}),
and time reversal, Eq. (\ref{105}), one obtains that the matrix elements
of $j_x$ and $j^{\pm}$ are real, while  $j_y$ is immaginary, as also
follows from Eq. (\ref{102}).

\section{Applications to deep inelastic scattering}
\label{S9}

 Let us first consider the problem of calculating the tensor (\ref{01}).
As follows from Eq. (\ref{31})

\begin{eqnarray}
&& W^{\mu\nu}=\frac{1}{4\pi}\sum (2\pi)^4\delta^{(4)} (P'+q-P")
\langle P',\chi'|J^{\mu}(0) |P",\chi"\rangle \cdot\nonumber\\
&& \langle P",\chi"|J^{\nu}(0) |P',\chi'\rangle
\label{85}
\end{eqnarray}
where the sum is taken over all possible final states with
four-momentum $P"$ and internal wave functions $\chi"$.

 We will consider the process in the reference frame where
${\bm P}_{\bot}'={\bm q}_{\bot}=0$, $P'_{z}=-P"_{z}=K>0$. Then,
as follows from Eqs. (\ref{81}),

\begin{eqnarray}
&& W^{\mu\nu}=\frac{1}{4\pi}\sum (2\pi)^4\delta^{(4)} (P'+q-P")
\langle \chi'|j^{\mu}(K\bm{e}_z;m,M") |\chi"\rangle \cdot\nonumber\\
&& \langle \chi"|j^{\nu}(-K\bm{e}_z;M",m) |\chi'\rangle
\label{86}
\end{eqnarray}
where $m$ is the mass of the nucleon and $M"$ is the mass of the
final state. As
follows from the delta function in Eq. (\ref{86}) and the definition of the
Bjorken variable $x=Q^2/2(P'q)$, in the Bjorken limit (when $Q\gg m$ and
$x$ is not too close to 0 or 1)
\begin{equation}
M^{"2}=\frac{Q^2(1-x)}{x},\quad K=\frac{Q}{2[(2-x)x]^{1/2}}
\label{87}
\end{equation}
(note that in the Bjorken limit these quantities do not depend on $m$).

 Using these expressions it is easy to show that in the reference
frame under consideration
\begin{equation}
P^{'+}=\sqrt{2} K,\quad P^{"+}=\sqrt{2} K(1-x),\quad
P^{"-}=\sqrt{2} K(2-x)
\label{88}
\end{equation}

Because of Eq. (\ref{50}), we have to choose the operators
$j^{\mu}(-K\bm{e}_z;M",m)$  and $j^{\mu}(K\bm{e}_z;m,M")$ in such a
way that
\begin{equation}
j^{\mu}(-K\bm{e}_z;M",m)^*=j^{\mu}(K\bm{e}_z;m,M")
\label{91}
\end{equation}
This expression makes it possible to rewrite Eq. (\ref{86}) in the form
\begin{eqnarray}
W^{\mu\nu}=\frac{1}{4\pi}\sum (2\pi)^4\delta^{(4)} (P'+q-P")
|\langle \chi'|j^{\mu}(K\bm{e}_z;m,M") |\chi"\rangle|^2
\label{92}
\end{eqnarray}

From Eq. (\ref{42}) one has
\begin{equation}
j^{\mu}(K\bm{e}_z;m,M")=\Pi J^{\mu}(0,P^{+};0,P^{"+})\Pi"
\label{89}
\end{equation}
where $\Pi'$ and $\Pi"$ are the orthogonal projectors onto the
states with the masses $m$ and $M"$ respectively, and
$P^{'+}$ and $P^{"+}$ are given by Eq. (\ref{88}).

A usual assumption in the parton model (where the final state
interaction is neglected) is that the current operator
can be taken in IA, viz.
 \begin{equation}
j^{\mu}(K\bm{e}_z;m,M")=\Pi J_{free}^{\mu}(0,P^{+};0,P^{"+})\Pi"
\label{90}
\end{equation}
where   $J^{\mu}(0)_{free}=\sum_{i=1}^N j_{free,i}^{\mu}$. In Sect.
\ref{S6}, we have already shown that, in general,  the free current
fulfills the extended Lorentz
covariance; moreover, as already noted, the Hermiticity property, Eq.
(\ref{60}), does not impose any further constraint in DIS, since
$m \not = M"$.
In the actual calculations of the structure functions, for any value of the
 momentum transfer, only three components of the current are needed
and can be chosen unconstrained with respect to the current
conservation,  while the fourth component can be determined through
the current conservation, see Eq. (\ref{67}). Therefore the structure
functions could be calculated by using the $+$ and ${\perp}$ components
of the free current operator in the Breit frame, even in the case
where the  final state interaction is present (cf., e.g., \cite{LPS1}).
However, in the parton model  all components
of the free current are compatible with current conservation in the
Bjorken limit. As a matter of fact, the $+$ and $\bot$
components of the vector $P$ are the same as for the system of free
quarks and gluons, and we have to discuss only the value of $P^{-}$, that
differs from the free one, for demonstrating that the free current
satisfies the current conservation in the parton model. For the initial
state, given
our choice of the reference frame, the value of $P^{'-}$ is negligible
and therefore the difference with respect to the value
in the free case is vanishing in the Bjorken limit. For the final state, if
 the interaction is disregarded,  as in the parton model,
${P"}^-$ is the same as for the free system. Therefore  Eq.
(\ref{67}) is satisfied by the free current in the Bjorken limit. The above
discussion
indicates that a consistent calculation of the tensor (\ref{01}) can be
achieved in the parton model by
replacing the full current operator $J^{\mu}(x)$  with the IA one. Obviously
this assumption does not imply that the IA current operator can be
adopted  for
calculating matrix elements in any other reference frames, apart the ones
reachable by front-form boosts.

 Summarizing, the
parton model does not contradict Poincar\'e covariance and current
conservation, although  the nucleon is described as a
bound state of quarks and gluons (see the discussion in Sect. \ref{S1}).
However, it is clear that the conditions
(\ref{56}), (\ref{72}), (\ref{73}), (\ref{63}), and (\ref{70a})
are not too restrictive and they allow many choices of $j^{\mu}$.

If the cluster separability is important, then in order to recover the
parton model results
(cf. \cite{lev3}) one can choose the em
current as follows
\begin{equation}
j^{\mu}(K\bm{e}_z;m,M")=\Pi' A J_{free}^{\mu}(0,P^{'+};0,P^{"+})A^{-1}\Pi"
\label{93}
\end{equation}
where $A$ is the packing operator.

\section{Elastic and inelastic scattering }
\label{S10}

\indent As  shown in Sect. \ref{S6},   in the case of elastic scattering
the Hermiticity condition   represents a  constraint to be
imposed  on  $j^{\mu}(K\bm{e}_z;M_i,M_i)$, besides the extended
Lorentz covariance and current conservation. Indeed, in this case
 the operators $j^{\mu}(K\bm{e}_z;M_i,M_i)$ and
$j^{\mu}(K\bm{e}_z;M_i,M_i)^*$
must be connected by $\pi$ rotations around the $x$ or $y$ axes (see, e.g.,
Eq. (\ref{60})). However, as already noted, Hermiticity can represent a non
trivial constraint also in the inelastic case, if the current operator is
defined for any value of the masses.

 Let $M_{i(j)}$ be the mass of a bound state $|\chi_{i(j)} \rangle$ of spin
 $S_{i(j)}$ and third component $S_{iz(jz)}$, $\Pi_{i(j)}$ be the projector
onto the corresponding subspace and
${\cal{J}}^{\mu}(K\bm{e}_z;M_i,M_j)$ be a current which fulfills
Eq. (\ref{56}) for any rotation $u_z$ around the $z$ axis (in the em
case we assume that ${\cal{J}}^{\mu}(K\bm{e}_z;M_i,M_j)$ satisfies also
Eqs. (\ref{72}) and (\ref{73})). As we have already shown, a possible
choice is the following one
\begin{eqnarray}
&& {\cal{J}}^{\mu}(K\bm{e}_z;M_i,M_j)=\Pi_i ~
J_{free}^{\mu}(0,P^{'+};0,P^{"+})~\Pi_j
\label{95}
\end{eqnarray}
where
\begin{equation}
P^{'+}=\frac{1}{\sqrt{2}}[(m^2+K^2)^{1/2}+K],\quad
P^{"+}=\frac{1}{\sqrt{2}}[(m^2+K^2)^{1/2}-K]
\label{96}
\end{equation}
and $K=Q/2$.
 Then a choice for the current compatible with the Hermiticity
condition, Eq. (\ref{60}), and with the extended Lorentz covariance is
\begin{eqnarray}
&&j^{\mu}(K\bm{e}_z;M_i,M_j)=\frac{1}{2}\{{\cal{J}}^{\mu}
(K\bm{e}_z;M_i,M_j)+\nonumber\\
&&{\cal{J}}^{\mu}(-K\bm{e}_z;M_j,M_i)^* \}
\label{97}
\end{eqnarray}
where ${\cal{J}}^{\mu}(-K\bm{e}_z;M_j,M_i)$ is given by a $\pi$ rotation of
${\cal{J}}^{\mu}(K\bm{e}_z;M_j,M_i)$ around the $x$ axis, in agreement with
Eq. (\ref{57}), viz.
\begin{eqnarray}
 && {\cal{J}}^{\mu}(-K\bm{e}_z;M_j,M_i) =
L^{\mu}_{\nu}[r_x(-\pi)]~exp(\imath \pi S_x)\nonumber\\
&&{\cal{J}}^{\nu} (K\bm{e}_z;M_j,M_i)
exp(-\imath \pi S_x)
\label{95'}
\end{eqnarray}
It is easy to show that the Hermiticity condition, Eq. (\ref{60}),
is satisfied by the current defined by Eqs. (\ref{97}) and (\ref{95'})
by noting that $exp(\imath 2\pi S_x)j^{\mu}(K\bm{e}_z;M_i,M_j)
exp(-\imath 2\pi S_x)=j^{\mu}(K\bm{e}_z;M_i,M_j)$.

It is also straightforward to check that such a  current fulfills
extended Lorentz covariance (i.e., Eqs. (\ref{56}), (\ref{72}) and
(\ref{73})). In particular, Eq. (\ref{56}) holds since $L[r_x(\pi)]
L[u_z]L[r_x(-\pi)]=L[-u_z]$ and $exp(-\imath \pi S_x) S_z
exp(\imath \pi S_x)=-S_z$. Parity covariance, Eq. (\ref{72}), holds
since $L[r_x(\pi)] \Lambda_y L[r_x(-\pi)]=\Lambda_y$ and
$U_{y~int}exp(\imath \pi S_x)=exp(-\imath 2\pi S_z) exp(\imath \pi S_x)
U_{y~int}$.

For the time reversal one can follow analogous arguments.

The Hermiticity condition, Eq. (\ref{60}), imposes the following
constraints on the matrix elements
\begin{eqnarray}
 && \langle  S_{jz} S_j  M_j |j^{\mp}(K\bm{e}_z;M_i,M_j)^*|M_i S_i S_{iz}
 \rangle = \delta_{S_{jz},S_{iz}}~(-1)^{(S_i-S_j)} \nonumber\\
 && \langle  -S_{jz} S_j  M_j |j^{\pm}(K\bm{e}_z;M_j,M_i)|M_i S_i -S_{iz}
  \rangle ,\nonumber \\
\label{103a}
\\
&& \langle  S_{jz} S_j  M_j |j_x(K\bm{e}_z;M_i,M_j)^*| M_i S_i S_{iz}
\rangle =
 (-1)^{(S_i-S_j)} \nonumber\\
&&\langle  -S_{jz} S_j  M_j |j_x(K\bm{e}_z;M_j,M_i)| M_i
 S_i -S_{iz} \rangle ,\nonumber \\
 \label{103b}
 \\
 && \langle  S_{jz} S_j  M_j |j_y(K\bm{e}_z;M_i,M_j)^*| M_i S_i S_{iz}
 \rangle =~-(-1)^{(S_i-S_j)}\nonumber\\
 && \langle  -S_{jz} S_j  M_j |j_y(K\bm{e}_z;M_j,M_i)| M_i S_i -S_{iz} \rangle
  \nonumber \\
 \label{103c}
\end{eqnarray}
since $\langle  S_z S  m |D^S[r_x(-\pi)]| m S S'_z \rangle=
\delta_{S_z,-S'_z}~(-1)^{-S+S_z} exp(\imath \pi S_z)$.
Equations (\ref{103b})-(\ref{103c}) are obviously compatible with
Eq. (\ref{102}).

Let us discuss now the current conservation in the elastic case
($M_i=M_j=m$;  $S_i=S_j=S$).   If
$j^+(K\bm{e}_z;m,m)$ and ${\bm j}_{\bot}(K\bm{e}_z;m,m)$ are
chosen as independent components of the operator
$j^{\mu}(K\bm{e}_z;m,m)$, then, in order to satisfy the continuity
equation, $j^-(K\bm{e}_z;m,m)$ has to be defined through Eq.
(\ref{67}), which reads for the elastic scattering
\begin{equation}
j^-(K\bm{e}_z;m,m)=j^+(K\bm{e}_z;m,m)
\label{minus}
\end{equation}
This condition implies that $j_z(K\bm{e}_z;m,m)=0$.

It is important to notice that Eq. (\ref{minus}) can be obtained as a
consequence of extended Lorentz covariance and Hermiticity, or in
other words that, in the elastic case, the extended Lorentz covariance
together with Hermiticity imposes current conservation. Indeed from
Lorentz covariance (Eq. (\ref{100})) only the diagonal matrix elements
of  $j^{\pm}(K\bm{e}_z;m,m)$ are different from zero, while from parity
and time reversal covariance we know that they are real. Then from
Hermiticity (Eq. (\ref{103a}))
\begin{eqnarray}
&&\langle m S S_z| j^-(K\bm{e}_z;m,m)^* | m S S_z \rangle =
\langle m S S_z| j^-(K\bm{e}_z;m,m)| m S S_z \rangle = \nonumber \\
&&\langle m S -S_z| j^+(K\bm{e}_z;m,m) | m S -S_z \rangle.
\label{herm}
\end{eqnarray}
Therefore using parity covariance (Eq. (\ref{104a})) we obtain
\begin{equation}
\langle m S S_z| j^-(K\bm{e}_z;m,m) | m S S_z \rangle =
 \langle m S S_z| j^+(K\bm{e}_z;m,m) | m S S_z \rangle.
\label{minus'}
\end{equation}
i.e., Eq. (\ref{minus}).

In particular the current defined by Eqs. (\ref{95}) and (\ref{97})
represents a possible choice. In this case it can be immediately seen that

\begin{eqnarray}
&&\langle m S S_z| j^+(K\bm{e}_z;m,m) | m S S_z \rangle =
 \langle m S S_z| {\cal J}^+(K\bm{e}_z;m,m) | m S S_z \rangle,
 \label{106} \\
&& \langle m S S_z| j_x(K\bm{e}_z;m,m) | m S S'_z \rangle = {1 \over 2}
[ \langle m S S_z| {\cal J}_x(K\bm{e}_z;m,m)
| m S S'_z \rangle -  \nonumber\\
&&\langle m S S'_z| {\cal J}_x(K\bm{e}_z;m,m) | m S S_z \rangle ]
\end{eqnarray}

An elementary application is represented by the elastic
scattering from a target with $S=0$. In this case only the  matrix
elements of $j^+$ are relevant; they are diagonal and real, and
therefore  the terms in Eq. (\ref{95}) are equal each other
(note that: i) $L^+_+[r_x(-\pi)]=1$, and ii) the matrix elements of
$exp(-\imath \pi S_x)$ are equal to one).

It is clear from the above discussion that in the elastic case one has (as
usual) only $2S+1$  independent matrix elements
for the em current defined in Eqs. (\ref{97}) and (\ref{95'}),
corresponding to $2S+1$ elastic form factors. The matrix elements of
$j_y$ and $j^-$ can be obtained from the matrix elements of  $j_x$ and
$j^+$, respectively, because of Eqs. (\ref{102}) and  (\ref{minus}).  As
follows from Eqs. (\ref{104a}) or (\ref{105}), there are $[S+1]$
non-zero independent matrix elements of $j^+$ ($[S+1]$ is the integer
part of $S+1$) and they can be chosen as the diagonal
ones with $S_z\geq 0$. The independent matrix elements of $j_x$  are
$[S+1/2]$ and can be chosen to be $\langle m S S_z| j_x(K\bm{e}_z;m,m) | m
S S_z-1 \rangle$ with $S_z\geq + 1/2$, as follows from Eqs. (\ref{103b}),
and (\ref{104b}) or (\ref{105}).

Also in the inelastic case, the matrix elements of
$j_y$ and $j^-$ can be obtained from the matrix elements of  $j_x$ and
$j^+$, respectively, because of Eqs. (\ref{102}) and  (\ref{67}). As
follows from Eqs. (\ref{104a}) or (\ref{105}), there are $[S+1]$
non-zero independent matrix elements of $j^+$ (with $S=min(S_i,S_j)$). The
matrix elements of $j_x$ are constrained by the rules $|S_{iz}-S_{jz}|=1$
and $S_{iz}\geq 0$ (see Eq. (\ref{104b})). For instance, in the case of the
transition from the nucleon  to a resonance  with $S_j = 1/2$ one has only
two   independent matrix elements, while for a transition to any resonance
with $S_j > 1/2$ one remains with only three independent matrix elements,
one for $j^+$ ($\langle 1/2|j^+|1/2\rangle $) and two for
$j_x$ ($\langle 1/2|j_x|-1/2\rangle $ and $\langle 1/2|j_x|3/2\rangle $),
corresponding to the three helicity amplitudes.

Summarizing, if one adopts the current defined by the Eqs. (\ref{95}),
(\ref{96}) and  (\ref{95'}) the extraction of  em form factors in both the
elastic and inelastic scattering is no more plagued by the ambiguities
discussed in the introduction.

\section{ Conclusion}
\label{S11}

 The results of the present paper show that the constraints
imposed by (extended) Poincar\'e covariance and current conservation
allows one to determine the current operator through  some auxiliary
operators
$j^{\mu}(K{\bm e}_z;M_i,M_j)$,
 which act only through internal variables and are covariant for rotations 
 around the $z$ axis. Unfortunately the  latter
 operators are not unique. This is in agreement with the results of
 Ref. \cite{PK} where it
has been
shown that all matrix elements of the current operator can be expressed in
terms of some set of fully unconstrained matrix elements.
We have demonstrated that it is possible to choose explicit models for
$j^{\mu}(K{\bm e}_z;M_i,M_j)$ such that all the necessary
requirements are satisfied. In particular, as noted in
Sects. \ref{S6}, \ref{S7}, \ref{S9} and \ref{S10} (see especially
Eqs. (\ref{90}), (\ref{93}) , (\ref{95}) and (\ref{97})), the operator
$j^{\mu}(K{\bm e}_z;M_i,M_j)$ can be obtained by projecting the free
current operator onto the subspaces corresponding to definite eigenvalues
of the mass and spin operators.

 It is also worth noting that, although the choice of
the Breit frame where the initial and final momenta are directed
along the $z$ axis is rather convenient, analogous results for
constructing the full current operator from auxiliary ones can be
derived by choosing any frame obtained from the Breit one by Lorentz
transformations corresponding to the group ${\tilde B}$,
where ${\tilde B}$ is a subgroup of SL(2,C) such that
${\tilde b}\in {\tilde B}$ if ${\tilde b}_{12}=0$ (it is easy to see
that ${\tilde B}$ is obtained from $B$ by adding rotations around
the $z$ axis). This follows from the
fact that in the front-form dynamics the Lorentz transformations
corresponding to the group ${\tilde B}$ are kinematical. In particular,
we can use the plus and $\bot$ components of the free current operator
for constructing  the same components of the auxiliary operators in
all such
reference frames and then the matrix elements of the minus components
can be determined (if necessary) from the continuity equation.

  In Sect. \ref{S9} we have considered the application of our results to
DIS. It
has been shown that for calculating the matrix elements of the current
operator in the infinite momentum frame where the initial and final
momenta are directed
along the $z$ axis we indeed can use the free current operator and this
does not contradict the fact that the nucleon is a bound state of
quarks and gluons. At the same time it has been briefly mentioned that
problems with locality and cluster separability exist. These problems
will be considered elsewhere \cite{LPS}.

  In Sect. \ref{S10} we have applied our results to the
 elastic and inelastic scattering for particles with arbitrary spin.
In contrast
with the approaches discussed in the Introduction, we have no problem with
the angular condition, since our model current is in agreement with extended
Poincar\'e covariance and current conservation, by construction. Therefore
the number of independent matrix elements of the current is equal to the
number of  physical form factors.

Finally, it should be pointed out that our approach, based on the reduction of 
the whole complexity of the Poincar\'e covariance to the SU(2) symmetry can
 represent a simple framework where to investigate  the possible many-body 
 terms to 
be added to the free current, since they must obviously fulfill the rotational 
covariance condition of Eq. (\ref{56}).

Numerical calculations for the deuteron elastic form factors and 
applications to the hadron elastic and transition form factors are in 
progress.

\begin{center} {\it Acknowledgements} \end{center}

\begin{sloppypar}
 The authors are grateful to F.Coester, W.Klink and H.J.Weber for
useful discussion. The work of one of the authors (FML) was
supported in part by grant 96-02-16126a from the Russian Foundation
for Basic Research.
\end{sloppypar}

\begin{center} {\large\bf Appendix A} \end{center}

\setcounter{equation}{0}
\def\theequation{A.\arabic{equation}}

 In this appendix it will be shown that
 \begin{eqnarray}
 &&W^{-1}(l,L(l)^{-1}{P_i \over M_i})= W(u,{K_i \over M_i})
 \label{1a}
\end{eqnarray}
where $K_i=B(H_{ij})^{-1}P_i$ and $u \in SU(2)$ is given by Eq.
(\ref{51}), viz.

\begin{eqnarray}
&&u=\beta(g')^{-1} l^{-1}\beta(g)
\label{2a}
\end{eqnarray}
with $g'=L(l)^{-1}H_{ij}$ and $g=H_{ij}$. Using Eqs. (\ref{Wigner})
and (\ref{2a}), one has

\begin{eqnarray}
&&W^{-1}(l,L(l)^{-1}{P_i \over M_i})= \beta(L(l)^{-1}{P_i
\over M_i})^{-1}l^{-1}\beta({P_i \over M_i})
\nonumber \\
&&W(u,{K_i \over M_i})=\beta(L(u){K_i \over M_i})^{-1}
\beta(g')^{-1} l^{-1} \beta(H_{ij})\beta({K_i \over M_i})
\label{3a}
\end{eqnarray}
Using the group property of the boosts (Eq. (\ref{9}))  one has

\begin{eqnarray}
 \beta(H_{ij})\beta({K_i \over M_i})=\beta(L(\beta(H_{ij}){K_i
\over M_i})= \beta({P_i \over M_i})
\label{4a}
\end{eqnarray}
and
\begin{eqnarray}
 \beta(L(u){K_i \over M_i})^{-1} \beta(g')^{-1} = \beta(L(\beta(g'))
L(u){K_i \over M_i})^{-1}
\label{5a}
\end{eqnarray}
Finally the  multiplicative rule of the Poincar\'e group yields
\begin{eqnarray}
&&L(\beta(g'))L(u){K_i \over M_i}=L(\beta(g'))L(\beta(g'))^{-1}
L(l)^{-1} L(\beta(H_{ij})){K_i \over M_i}= \nonumber \\
&&L(l)^{-1} {P_i \over M_i}
\label{6a}
\end{eqnarray}
 Then, collecting the results from Eqs. (\ref{3a})-(\ref{6a})
we  find that Eq. (\ref{1a}) holds. The same it is true for
$W^{-1}(l,L(l)^{-1}{P'_j / M_j})= W(u,{K'_j / M_j})$
\end{sloppypar}
 
\end{document}